\documentclass[aps,showpacs]{revtex4}

\usepackage[dvips]{graphics,color}
\usepackage{graphicx}
\usepackage{dcolumn}
\usepackage{bm}
\usepackage{epsfig}

\newcommand{\sla}{\hskip-0.2cm{\slash}}

\newcommand{\beq}{\begin{equation}}
\newcommand{\eeq}{\end{equation}}
\newcommand{\beqa}{\begin{eqnarray}}
\newcommand{\eeqa}{\end{eqnarray}}
\newcommand{\bd}[1]{ \mbox{\boldmath $#1$}}

\begin{document}
\def\ii{\'\i}

\markboth{Peter O. Hess and Walter Greiner}
{PSEUDO-COMPLEX FIELD THEORY}

\title{
PSEUDO-COMPLEX FIELD THEORY
}

\author{Peter O. Hess$^{1,2}$ and Walter Greiner$^{1}$}
\affiliation{
$^1$Frankfurt Institute of Advanced Studies, Johann Wolfgang Goethe Universit\"at,
Max-von-Laue Str. 1, 60438 Frankfurt am Main, Germany \\
$^2$Instituto de Ciencias Nucleares, UNAM, Circuito Exterior, C.U., A.P. 70-543, 04510 M\'exico D.F., Mexico
}

\begin{abstract}
{
A new formulation of field theory is presented, based on a
pseudo-complex description. An extended group structure
is introduced, implying a minimal scalar length, rendering the theory
regularized a la Pauli-Villars.
Cross sections
are calculated for
the scattering of an electron at an external Coulomb field and
the Compton scattering. Deviations due to a smallest
scalar length are determined.
The theory also permits a modification
of the minimal coupling scheme, resulting in a generalized
dispersion relation. A shift of the Greisen-Zatsepin-Kuzmin-limit
(GZK) of the cosmic ray spectrum is the consequence.
}
\end{abstract}

\maketitle

\section{Introduction}

The (Quantum) Field Theory (FT, in general) is a great success story of physics.
For example, Quantum Electrodynamics \cite{greiner} describes many
physical processes to an unprecedent precision, rarely known in
science. The Lorentz symmetry plays an important r\^{o}le, which
allows to formulate several conservation laws, like the PCT theorem.
However, experiments on the cosmic ray spectrum \cite{agasa}
put some doubts on this symmetry.
Further high energy events were measured in Ref. \cite{fly,haverah,yakutsk}.
Under the assumption of Lorentz
symmetry, a maximal possible energy at 10$^{20}$~eV is predicted, called the
Greisen-Zatsepin-Kuzmin (GZK) limit (or cutoff)
\cite{gkz1,gkz2}. It originates from the interaction
of high energy protons with the photons of the Cosmic-Microwave-Background
(CMB). At very large energies, corresponding to the GZK limit, the
energy of the photon in the eigen-system of the proton is large enough
to create pions. This process subtracts energy from the proton and, thus,
no higher energy protons can reach earth.
Of course, the assumption is important that these hight energy particles
are produced very far from earth, for example,
in the early universe.
In a recent article \cite{HiRes}, reporting on the results of
the HiRes experiment, the GZK cutoff was claimed to be observed.
The results of AGASA and HiRes are put on trial in
the Pierre-Auger experiment \cite{auger,lukas}, where final data
have not been published yet.

In spite of the great success of FT, the occurrences of ultraviolet
divergences is troublesome, requiring intelligent
subtractions of infinities. These are due to the assumption of permitting
arbitrary large momenta, i.e. small lengths. However, physics might
change at the order of the Planck length or even before. Adding
a smallest length ($l$) corresponds to introduce a momentum cutoff, which
eliminates the infinities, though, renormalization of physical
parameters, like the charge or mass, have still to be applied.
A smallest length scale $l$ must have an influence on the position
of the GZK cutoff. Conversely, if a shift is observed it will put
a value to the $l$. As we will see in this contribution, the effect of
a smallest length is larger for high energy processes. The atomic energy scale
are too small. Investigating the GZK limit gives a good opportunity
to look for a smallest length scale.

Up to now, the main efforts to explain the non-existence of the GZK limit
concentrates on the violation of Lorentz symmetry.
In accordance with our last observation, a minimal length is
introduced in most models (see, e.g., the {\it Double Special
Relativity} \cite{dsr1,dsr2} or {\it spin-networks} \cite{smolin}).
The length is handled as a vector,
subject to Lorentz contraction. Another way to break Lorentz invariance
is to assign a velocity with respect to an universal frame, breaking
rotational and, thus, also Lorentz symmetry. This is proposed
in \cite{goenner1,goenner2,goenner3}, based on a geometric approach.
In \cite{bertolami} a Lorentz breaking interaction was considered,
also containing a preferred oriented vector. In \cite{coleman1,coleman2}
Lorentz breaking interactions in the Lagrange density were investigated
too on a general basis.

In \cite{hess1} an alternative is proposed, extending the
Lorentz group to a larger one. The formalism is based on a
pseudo-complex extension of the Lorentz group \cite{schuller2}, where
pseudo-complex numbers have to be introduced, also called
hyperbolic or hypercomplex. Large part of the mathematics is described
in detail in Ref. \cite{crumeyrolle}. It also implies to formulate a
pseudo-complex
version of the field theory, which is proposed schematically in
\cite{schuller3}, however, without any calculations of physical processes.
Adding a term to the Lagragian, which simulates that the
interaction happens within a finite size
of space-time and not at a point (due to the occurrence of a minimal length
scale $l$), changes the dispersion relation \cite{hess1}.
The minimal length scale ($l$)
enters and modifies the dispersion relation, giving rise to
a shift in the GZK limit. However, the maximal predicted cutoff,
under reasonable assumptions and independent on the choice
of further structure of the additional interaction term,
is by only a factor 2.
The difference is proportional to $l^2$ and increases with
energy. The GZK cutoff gives us the opportunity of investigating
such high energy events. If not observed, at least we can obtain
an upper limit on the smallest lengthy scale $l$.

Consequently, the change in the dispersion relation is visible only
at high energies, comparable to the GZK scale.
At low energies, the dispersion relation is to very high
approximation maintained. One may ask however, if the smallest
length $l$ may also produce deviations at intermediate energies,
for example,
in the TeV range, accessible to experiment now. In order to be
measurable, we look for differences in
the cross section of a particular reaction, of the lowest power in
$l$ possible.

The advantage of the proposed extended field theory is obvious: All
symmetries are maintained and, thus, it permits the calculation of
cross sections as we are used to. Still, an invariant length scale appears,
rendering the theory regularized and reflecting the deviation
of the space-time structure at distances of the order of the
Planck length.

The main objective of this paper is to formulate the
pseudo-complex extension of the standard field theory (SFT).
For the extension we propose the name {\it Pseudo-Complex Field Theory}
(PCFT). First results are reported in \cite{hess1}.

The structure of the paper is as follows: In section 2 the pseudo-complex
numbers are introduced and it is shown how to perform calculations, like differentiation
and integration. This section serves as a quick reference guide to the reader
unfamiliar with the concept of pseudo-complex numbers.
In section 3 the pseudo-complex Lorentz and Poincar\'e groups
are discussed.
The representations of the pseudo-complex Poincar\'e group are indicated.
Section 4 introduces a modified variational procedure, required in
order to obtain a new theory and not two separated old ones. The
language is still classical. As examples, scalar and Dirac fields are
discussed and an extraction procedure, on how to obtain physical observables,
is constructed and at the end formally presented.
Section 5 is dedicated to the symmetry properties of the PCFT.
Finally, in section 6 the quantization formalism is proposed. In section
7 a couple of cross sections are calculated within the PCFT: i) The
dispersion of a charged particle at a Coulomb field and ii) the Compton scattering.
One could also consider high precision measurements, like the Lamb shift and the
magnetic moment of the electron. These, however,
require higher order Feynman diagrams, which will explode the scope
of the present paper. This will be investigated in a future article.
The language will be within Quantum Electrodynamics and effects from the
electro-weak unification will be discarded, for the moment.
In section 8 we will show some relations to geometric approaches,
which also contain a scalar length parameter.
The results of this section will give important implications for
the topic treated in Section 9, where the theory is extended such
that the GZK limit is shifted.
Finally, section 10 contains the conclusions and an outlook.

The paper contains at the beginning
an explanatory part of a work already published
\cite{schuller2,crumeyrolle,schuller3,schuller1,schuller4},
however, in a quite dense form.
Parts, published in \cite{schuller2,schuller3,schuller1,schuller4},
had to be revised and inconsistencies, physical and mathematical ones,
were corrected.
It also contains new contributions to
the pseudo-complex formulation.
The main motivation is to make this contribution self-contained and
to expand the very short presentations, given in several
different contributions
of the pseudo-complex formulation,
such that the reader appreciates the global context.
The new and additional contributions can be found in the mathematical
part, to the
representation theory, how to extract physical observables (like
cross sections), the quantization procedure and the calculation of cross
sections.

\section{Pseudo-Complex Numbers and derivatives}

The pseudo-complex numbers, also known as
{\it hyperbolic} \cite{crumeyrolle} or
{\it hypercomplex} \cite{kantor},
are {\it defined} via

\beqa
X & = & x_1 + I x_2 ~~~,
\eeqa
with $I^2=1$. This is similar to the common complex notation except for
the different behavior of $I$. An alternative presentation is to introduce

\beqa
\sigma_\pm & = & \frac{1}{2} \left( 1 \pm I \right)  \nonumber \\
\eeqa
with
\beqa
\sigma_\pm^2 & = & 1 ~~~, \sigma_+\sigma_- = 0 ~~~.
\eeqa
The $\sigma_\pm$ form a {\it zero divisor basis}, with the zero divisor
defined by
$\bd{P}^0 = \bd{P}^0_+ \cup \bd{P}^0_-$, with
$\bd{P}^0_\pm=\left\{ X=\lambda \sigma_\pm| \lambda ~\epsilon~ \bd{R} \right\}$.

This basis is used to rewrite the pseudo-complex numbers as

\beqa
X & = & X_+ \sigma_+ + X_- \sigma_- ~~~,
\eeqa
with

\beqa
X_\pm = x_1 \pm x_2  ~~~.
\eeqa
The set of pseudo-complex numbers is denoted by
${\bd P} = \left\{ X=x_1+Ix_2 | x_1,x_2\epsilon {\bd R} \right\}$.

The pseudo-complex conjugate of a pseudo-complex number is 

\beqa
X^* & = & x_1 - I x_2 = X_+\sigma_- + X_- \sigma_+ ~~~.
\eeqa
We use the notation with a star for the pseudo-complex conjugate
and a bar ($\bar{X}$) to denote the usual complex conjugate, i.e, the
pseudo-real and pseudo-imaginary part can also be complex, though,
in this section we assume that they are real for didactical reasons.
The {\it norm} square of a pseudo-complex number is given by

\beqa
|X|^2 = XX^ * & = & x_1^2 - x_2^2 ~~~.
\eeqa
There are three different possibilities:

\beqa
x_1^2 - x_2^2  & > & 0 ~~~,~~~ "{\rm space~like}" \nonumber \\ 
x_1^2 - x_2^2  & < & 0 ~~~,~~~ "{\rm time~like}" \nonumber \\ 
x_1^2 - x_2^2  & = & 0 ~~~,~~~ "{\rm light~cone}" ~~~,
\eeqa
where the notation in "..." stays for the analogy to the structure of the
1+1-dimensional Minkowski space. In each subsection, a different
parametrization of the pseudo-complex sector can be applied
\cite{schuller2,schuller0}.

\noindent
i) Positive norm:

The  presentation of a pseudo-complex number is very analogous to the
usual complex one

\beqa
X & = & R e^{I\phi}  =  R({\rm cosh}(\phi ) + I {\rm sinh}(\phi ) ) \nonumber \\
\eeqa
with
\beqa
x_1 & = & R {\rm cosh}(\phi ) ~~~,~~~ x_2 = R {\rm sinh}(\phi ) ~~~.
\eeqa
The inverse relation is given by

\beqa
R & = & \pm\sqrt{x_1^2 - x_2^2} \nonumber \\
{\rm tanh}(\phi ) & = & \frac{x_2}{x_1} ~~~.
\eeqa

There are two cases: $R>0$ and $R<0$, corresponding to the "right" and
"left" cone, respectively. 
Constant $R$ corresponds to hyperboloids either
on the right or left cone.

\noindent
ii) Negative norm:

The only difference is an additional $I$ in the parametrization of
the pseudo-complex number, i.e.,

\beqa
X & = & R Ie^{I\phi}  =  R(I{\rm cosh}(\phi ) + {\rm sinh}(\phi ) ) \nonumber \\
\eeqa
with
\beqa
x_2 & = & R {\rm cosh}(\phi ) ~~~,~~~ x_1 = R {\rm sinh}(\phi ) ~~~.
\eeqa
The inverse transformation is

\beqa
R & = & \pm\sqrt{x_2^2 - x_1^2} \nonumber \\
{\rm tanh}(\phi ) & = & \frac{x_1}{x_2} ~~~.
\eeqa

There are two cases: $R>0$ and $R<0$, corresponding to the "upper" and
"lower" cone, respectively.
Constant $R$ corresponds to either hyperboloids
on the upper or lower cone.

\noindent
iii) Zero norm:

The parametrization is given by

\beqa
X & = & \lambda \frac{1}{2}\left( 1 \pm I \right) = \lambda \sigma_\pm
\eeqa
With $X^*X=0$ it satisfies the condition for the zero norm.

In the $(x_1,x_2)$ plane, this subspace is represented by diagonal
lines, which depict the zero divisor branch.

\begin{figure}[t]
\centerline{\epsfxsize=7cm\epsffile{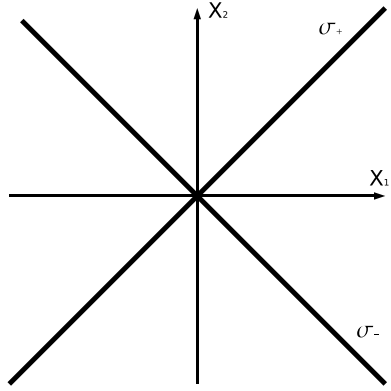}}
\vskip -2cm
\caption{
Illustration of the pseudo-complex plane for the variable
$X=X_1 + IX_2$ = $X_+ \sigma_+ + X_- \sigma_-$. The horizontal and
vertical line correspond to the pseudo-real and pseudo-imaginary
axes, respectively. The diagonal lines represent the zero divisor branch.
}
\label{fig1}
\end{figure}

The different sectors are illustrated in Fig. \ref{fig1}.

\vskip 0.5cm
As can be seen, the structure of the space is very similar to the one
of the Minkowski space. In fact, the structure corresponds to the group
$O(1,1)$.

\vskip 0.5cm

A useful rule of any function $F(X)$, which is expanded into
a Taylor series, can be written as

\beqa
F(X) & = & F(X_+) \sigma_+ + F(X_-) \sigma_-  
\eeqa
and a product of two functions $F(X)$ and $G(X)$ satisfies

\beqa
F(X)G(X) & = & F(X_+)G(X_+) \sigma_+ + F(X_-)G(X_-) \sigma_-  ~~~.
\eeqa
This is proved, using $\sigma_\pm^2=1$ and $\sigma_+\sigma_-=0$
and

\beqa
X^n & = & (X_+ \sigma_+ + X_- \sigma_-)^n \nonumber \\
& = & X_+^n \sigma_+ + X_-^n \sigma_- ~~~,
\eeqa
for arbitrary $n$ (note, that $\sigma_\pm^n = \sigma_\pm$, for all $n$).
As an example, we have

\beqa
e^{X} & = & e^{X_+} \sigma_+ + e^{X_-} \sigma_- ~~~.
\eeqa

\subsection{Differentiation}

A function $f(X)=f_1(X)+If_2(X)$
is called pseudo-complex differentiable if it fulfills
the {\it pseudo-}Cauchy-Riemann equations

\beqa
\partial_1f_1 & = & \partial_2f_2 \nonumber \\
\partial_2f_1 & = & \partial_1f_2 ~~~,
\eeqa
with $\partial_k=\frac{\partial}{\partial x_k}$.
This definition of a derivative is completely analogous to the one
with the usual complex numbers (see, e.g., \cite{peschl}).
It leads to the following expression for the pseudo-complex
derivative:

\beqa
\frac{D}{DX} & = & \frac{1}{2}(\partial_1 + I\partial_2) \nonumber \\
& = & \frac{1}{2} \left[ (\partial_1 + \partial_2) \sigma_+
+ (\partial_1 - \partial_2) \sigma_- \right] \nonumber \\
& = & \partial_+ \sigma_+ + \partial_- \sigma_-
~~~.
\eeqa
Care has to be taken with the zero divisor branch
$\bd{P}^0$ (see definition above).
Pseudo-complex derivatives are only defined outside this branch, leading
to a separation between areas of different norm. Functions can, therefore,
only be expanded in a Taylor series within each sector.

Using the analogy to the usual complex numbers, we could write $dX$
instead of $DX$, etc., keeping in mind that we deal with a pseudo-complex
derivative. Nevertheless, for the moment we keep this notation. All what
we have to remember is that the rules are similar, e.g.
$\frac{D(X^n)}{DX}=n X^{n-1}$.

A function in $X$ is called {\it pseudo-holomorph}
in $X$, when it is differentiable in a given area around $X$, just similar
to the definition of normal complex functions.

The extension to a
derivative with more than one dimension index is direct, i.e.,

\beqa
\frac{D}{DX^\mu}=\frac{1}{2}(\partial_{1,\mu} + I\partial_{2,\mu}) ~~~.
\eeqa

The derivative can also be extended to fields (in the sense as
described in any text book on Classical Mechanics discussing the
continuous limit \cite{greiner-cl}.
A functional derivative with respect to a pseudo-complex field
$\Phi_r = \phi_{1,r} +I \phi_{2,r}$ ($r=1,2,...$) is given by

\beqa
\frac{D}{D\Phi_r (X)}=\frac{1}{2}(\frac{\partial}{\partial \Phi_{1,r} (X)}
+ I\frac{\partial}{\partial \Phi_{2,r} (X)}) ~~~.
\eeqa
Similarly defined are functional derivatives with respect to
$D_\mu \Phi_r$. For example the derivative of $D_\nu\Phi (X) D^\nu\Phi (X)$
with respect to $D_\mu \Phi (X)$ gives

\beqa
\frac{D_\nu\Phi (X) D^\nu\Phi (X)}{D_\mu \Phi (X)} & = & 2 D^\mu\Phi (X) ~~~.
\eeqa

\subsection{Integration}

In general, we have to provide a {\it curve} $X(t)=x_1(t)+Ix_2(t)$,
with $t$ being the curve parameter,
along which we
would like to perform the integration. A pseudo-complex integral
can be calculated via real integrals (as for the normal complex case):

\beqa
\int F(X) dX & = & \int dt \left( \frac{dx_1}{dt} + I \frac{dx_2}{dt} \right)
F(X(t)) ~~~.
\eeqa

However, no residual theorem exists.
Thus, the structure of pseudo-complex
numbers is very similar to the usual complex ones but not completely,
due to the appearance of the zero divisor branch.
This reflects the less stringent algebraic structure, i.e., that the
pseudo-complex numbers are not a field but a ring.

\subsection{Pseudo-complex Fourier integrals}

In $d$-dimensions, the Fourier transform of a function $F(X)$ and its inverse 
can be defined via

\beqa
F(X) & = & \frac{1}{(2\pi )^{\frac{d}{2}}} \int d^dP \tilde{F}(P)
e^{i P\cdot X}
\nonumber  \\
\tilde{F}(P) & = & \frac{I^{n_2}}{(2\pi )^{\frac{d}{2}}} \int d^dX F(X)
e^{-i P\cdot X}
~~~,
\eeqa
with both $X$ and $P$ in general pseudo-complex, 
$P\cdot X$ = $(P^\mu X_\mu)$ and $n_2$ being the number of integrations
in the "time-like" sector. Here, we restrict to straight lines in either
the "space-like" or "time-like" sector. Straight lines
in the "space-like" sector (here, for example of the
coordinate $X$) are parametrized as:
$X = R exp(I\phi_0)$ ($\phi_0=const$). For the integration in the "time-like"
sector we have $X=IRexp(I\pi_0)$.

With this definition of the Fourier transform,
in 1-dimension, the $\delta$-function is given by

\beqa
\tilde{\delta} (X-Y) & = & \frac{1}{2\pi} \int dP e^{iP(X-Y)} \nonumber \\
& = &I^\xi \left( \delta (X_+-Y_+) \sigma_+ + \delta (X_--Y_-) \sigma_-
\right)
~~~,
\eeqa
with $\xi = 0$ if the integration along a straight line is performed in
the "space-like" sector and it is equal to 1 if the integration is
performed along a line in the "time-like" sector.
For a more detailed description, consult Appendix A.

\section{Pseudo-complex Lorentz and Poincar\'e groups}

Finite transformations in
the pseudo-complex extension of the Lorentz group are
given by $exp(i\omega_{\mu\nu}\Lambda^{\mu\nu})$, where
$\omega_{\mu\nu}$ is a pseudo-complex group parameter
$(\omega_{\mu\nu}^{(1)} + I\omega_{\mu\nu}^{(2)})$ and $\Lambda^{\mu\nu}$
are the generators \cite{gen-lo,mueller}.
The finite transformation with pseudo-complex parameters form the
pseudo-complex Lorentz group $SO_{\bd{P}}(1,3)$.

When acting on functions in the pseudo-complex coordinate
variable $X^\mu$, the
representation of the generators of the Lorentz group is,
using as the momentum operator $P^\mu = \frac{1}{i}D^\mu$
(=$P_1^\mu + I P_2^\mu$ = $P_+^\mu\sigma_+ + P_-^\mu\sigma_-$), in the
two possible representations:

\beqa
\Lambda^{\mu\nu} & = & X^\mu P^\mu - X^\nu P^\mu \nonumber \\
& = & \Lambda_+^{\mu\nu}\sigma_+ + \Lambda_-^{\mu\nu}\sigma_- ~~~,
\label{L-X}
\eeqa
with

\beqa
\Lambda_\pm^{\mu\nu} & = & X_\pm^\mu P_\pm^\mu - X_\pm^\nu P_\pm^\mu ~~~.
\eeqa

The pseudo-complex Poincar\'e group is generated by 

\beqa
P^\mu & = & iD^\mu = i \frac{D}{DX_\mu}= P^\mu_+ \sigma_+ + P^\mu_-\sigma_-
\nonumber \\ 
\Lambda^{\mu\nu} & = & \Lambda_+^{\mu\nu}\sigma_+ + \Lambda_-^{\mu\nu}\sigma_-  ~~~,
\eeqa
with $P^\mu_\pm=P^\mu_1\pm P^\mu_2$.

As before, finite transformations of the pseudo-complex Poincar\'e group
are given by $exp(i\omega_+\cdot L_+)\sigma_+$ +
$exp(i\omega_-\cdot L_-)\sigma_-$, with $(\omega_\pm\cdot L_\pm)$ =
$\omega_{\pm ,i} L_\pm^i$, $L_\pm^i$ being either
$\Lambda_\pm^{\mu\nu}$ or $P_\pm^\mu$, and in general distinct
pseudo-real group parameters
$\omega_{+,i}$ and $\omega_{-,i}$.
Only when $\omega_{-,i}=\omega_{+,i}$ the
group parameters $\omega_i$ are pseudo-real and standard Lorentz group is
recovered.

A Casimir of the pseudo-complex Poincar\'e group is

\beqa
P^2 & = & P_\mu P^\mu ~=~ \sigma_+ P^2_+ + \sigma_- P^2_- ~~~.
\eeqa
Its eigenvalue is $M^2 = \sigma_+ M_+^2 + \sigma_- M_-^2$, i.e.,
a pseudo-complex mass associated to each particle.
The Pauli-Ljubanski vector is given by \cite{gen-lo}

\beqa
W_\mu & = & -\frac{1}{2} \epsilon_{\mu\gamma\alpha\beta} P^\gamma
\Lambda^{\alpha\beta} \nonumber \\
& = & W_{\mu +} \sigma_+ + W_{\mu -} \sigma_-
~~~.
\label{ljubanski1}
\eeqa
The $\sigma_\pm$ parts of this vector are

\beqa
W_{\mu\pm} & = & -\frac{1}{2} \epsilon_{\mu\gamma\alpha\beta} P^\gamma_\pm
\Lambda^{\alpha\beta}_\pm ~~~.
\label{ljubanski2}
\eeqa
Thus, two mass scales are associated to a particle, namely $M_+$ and
$M_-$, which are {\it in general} different.
Its interpretation will be discussed below.

\subsection{Interpretation of a pseudo-complex transformation}

In this subsection the effect of a transformation
with pseudo-imaginary group parameters is revisited. The first
steps towards an extraction procedure, on how to obtain
physically observable numbers, are presented.
Step by step, this will be
complemented to a final extraction procedure.
Later, in section
4.4, all building blocks are united and a formal
justification will be given.

A finite transformation of the pseudo-complex Lorentz group
is expressed by $exp\left( i\omega_{\mu\nu} \Lambda^{\mu\nu}\right)$
= $exp\left( i \omega \cdot \Lambda \right)$, where
$\omega_{\mu\nu}$ is pseudo-complex.
In order to study the
effect of a pseudo-complex transformation, it suffices to restrict
to a purely pseudo-imaginary $\omega_{\mu\nu} \rightarrow I\omega_{\mu\nu}$,
where we extracted the $I$. Thus, a finite transformation
is given by

\beqa
\Lambda_{\mu}^{~\nu} & = & exp\left( iI\omega \cdot \Lambda\right) \nonumber \\
& = & \Lambda_{1,\mu}^{~~~\nu} + I \Lambda_{2,\mu}^{~~~\nu} ~~~,
\eeqa
were the transformation is divided into its pseudo-real and pseudo-imaginary
components. The pseudo-real part can again be associated to a standard Lorentz
transformation.

Now, let us consider a co-moving four-bein along the world-line of
the observer. The unit vectors are denoted by $\bd{e}_\mu$.
Applying to it the pseudo-complex transformation leads to new, now
pseudo-complex, unit vectors $\bd{E}_\mu$, which are related to the old ones
via

\beqa
\bd{E}_\mu & = \Lambda_{1,\mu}^{~~~\nu} \bd{e}_\nu + lI \Omega_\mu^{~~\nu} \bd{e}_\nu ~~~,
\eeqa
with

\beqa
\Omega_\mu^{~\nu} & = & \frac{1}{l}(\Lambda_1^{-1})_\mu^{~\lambda}
(\Lambda_2)_\lambda^{~\nu} ~~~.
\eeqa
It is straight forward to show that the following symmetry properties hold

\beqa
(\Lambda_1)_{\mu\nu} & = & (\Lambda_1)_{\nu\mu} \nonumber \\
\Omega_{\mu\nu} & = & - \Omega_{\nu\mu} ~~~.
\label{symm}
\eeqa

Let us consider as a particular transformation the boost in direction 1, the
presence of the $I$ requires a pseudo-imaginary angle $I\phi$ of the boost.
Using ${\rm cosh}(I\phi )={\rm cosh}(\phi )$ and
${\rm sinh}(I\phi )=I{\rm sinh}(\phi )$, the
transformation acquires the form

\beqa
\Lambda & = &
\left(
\begin{array}{cccc}
{\rm cosh}(\phi ) & I~{\rm sinh}(\phi ) & 0 & 0 \\
I~{\rm sinh}(\phi ) & {\rm cosh}(\phi ) & 0 & 0  \\
0 & 0 & 1 & 0 \\
0 & 0 & 0 & 1 \\
\end{array}
\right)         ~~~.
\label{lambda}
\eeqa
With Eq. (\ref{lambda}) this gives the relation for the relevant components

\beqa
\Omega_0^0 & = & \Omega_1^1 ~=~ 0 \nonumber \\
\Omega_0^1 & = & \Omega_1^0 ~=~ \frac{1}{l}{\rm tanh}(\phi ) ~~~,
\label{acc}
\eeqa
where the matrix element $\Omega_0^1$ describes the acceleration of the
particle.
Lowering the indices reproduces the required symmetry properties of
Eq. (\ref{symm}).
As a special case, consider $(\Lambda_1)_\mu^{~\nu} = \delta_\mu^\nu$.
The four-bein vectors acquire the form

\beqa
\bd{E}_\mu & = & \bd{e}_\mu + lI \Omega_\mu^\nu \bd{e}_\nu  \nonumber \\
& = & \bd{e}_\mu + lI \frac{d\bd{e}_\mu}{d\tau}\label{E-e}
~~~,
\eeqa
where $\tau$ is the eigen-time.
The pseudo-imaginary component describes the changes of the four-bein
vectors with time, i.e., for the 0-component it gives the acceleration,
described by the Frenet-Serret tensor $\Omega_\mu^\nu$,
of the co-moving system along the world line of the observer.
The form of $\Omega_\mu^\nu$ implies a maximal value
for the acceleration (see Eq. (\ref{acc})) which is, using the
limit ${\rm tanh}(\phi ) \rightarrow 1$ for $\phi \rightarrow \infty$,

\beqa
\Omega_0^1 ~=~ \frac{1}{l}{\rm tanh}(\phi ) & \leq & \frac{1}{l} ~~~.
\eeqa

In conclusion, {\it the pseudo-imaginary component of the group parameter
results in the appearance of a maximal acceleration and, thus, the theory
contains
a minimal length scale}. Which value this $l$ acquires, cannot be
decided yet.
We will see that it should be of the order of the Planck length.
Important to note is that $l$ is a {\it scalar} parameter which
is not subject to a Lorentz contraction.

When transformations of $\Lambda_{kl}$, with $k,l=1,2,3$ are considered,
the transformed systems correspond to rotating ones \cite{schuller0}.

Equation (\ref{E-e}) suggests to propose as a new coordinate

\beqa
X^\mu & = & x^\mu +lI\frac{dx^\mu}{d\tau} \nonumber \\
&& x^\mu + lI u^\mu ~~~,
\label{x-geom}
\eeqa
with $\tau$ the eigen-time and $u^\mu$ as the four-velocity
of the observer.
In general, the $X_1^\mu$ and $X_2^\mu$ in $X^\mu = X_1^\mu + I X_2^\mu$
are {\it linear independent}. Eq. (\ref{x-geom}) proposes to
fix the pseudo-imaginary component, when mapped to a physical system,
using this geometrical argument.

The pseudo-imaginary component of $P^\mu$ allows the following
interpretation:
When we apply a finite transformation $exp(ilIb_\mu P^\mu )$ to $X^\mu$,
which is pseudo-imaginary as in the former case, it only affects the
pseudo-imaginary component of $X^\mu$, namely $u^\mu$
$\rightarrow$ $l(u^\mu + b^\mu)$. This
action changes the four-velocity and, thus, corresponds to an acceleration.
Therefore, we can associate to the pseudo-imaginary part of the translation
operator an accelerations, too. 
This will play an important role in sections 8 and 9, when
a modified procedure is proposed on how to extract
physically observable numbers.
In general, the two components of
$P^\mu$ are linear independent and only at the end a choice
of the pseudo-imaginary component is applied.

As was shown, the pseudo-complex extension includes systems which
are accelerated (and rotated, when rotational transformations
with pseudo-imaginary group parameter are used). However,
when we want to determine physical observables, we have to do it in
an inertial frame because only there the vacuum is well defined.
This implies to select a subset of systems, corresponding to inertial
ones, with respect to which the comparison to SFT is possible.
Because the $P_2^\mu$ component is associated to acceleration,
it is suggested to set $P_2^\mu$ to zero.
However, adding to the Lagrange density a {\it fixed}, i.e. not
linear independent, pseudo-complex component to
$P^\mu$ may {\it simulate} the effect of acceleration during interaction.

For the moment, we will put the pseudo-imaginary component to zero, when
extracting physical results and only later, in section 9, we will explore
the consequences permitting a linear {\it dependent} pseudo-complex
component.

\subsection{Representations}

One implication of the above description for fields
$\Phi_r(X)$, with $r$ denoting internal degrees of freedom,
is that it depends on
the pseudo-complex coordinate $X$. In the zero-divisor basis this
field acquires the form $\Phi_{r,+}(X_+)\sigma_+ + \Phi_{r,-}(X_-)\sigma_-$.
The Casimir operator $W^2=W_\mu W^\mu$ of the Poincar\'e group
is proportional to $M^2J^2$ = $M_+^2J_+^2\sigma_+ + M_-^2J_-^2\sigma_+$,
with $M$ the pseudo-complex mass and
$J^2=J_+^2\sigma_+ + J_-^2\sigma_-$ the total spin squared.
Spin is a conserved quantity and the pseudo-complex fields have to be
eigenstates of this spin-operator. Because the eigenvalue is real,
the eigenvalues of $\Phi_{r,\pm}$ with respect to $J_\pm^2$
have to be the same.

The representation theory of the new field theory is completely
analogous to the standard procedure \cite{gen-lo} and it is not necessary
to elaborate on it further. The same holds for the Poincar\'e group.

The eigenvalue $M^2$ of $P^2$ is pseudo-complex and results in two
mass scales, namely $M_+$ and $M_-$. One of these scales will be associated
to the physical mass $m$, as will be shown further below. The other scale
will be related to $l^{-1}$, setting it equal to the Planck mass.

\section{Modification of the variational procedure}

Up to now, it seems that everything can be written in parallel, i.e.,
one part in $\sigma_+$ and the other one in $\sigma_-$. {\it In order to
obtain a new theory, both parts have to be connected.}

Given a Lagrange density, one is tempted to introduce in the pseudo-complex
space an action $S=\int d^dX~{\cal L}$. The equations of motion are
obtained by a variational procedure. However, if we
require that $\delta S=0$, we come just back to two {\it separated}
actions and two {\it separated} wave equations, because we can write
the action as $S=S_+\sigma_+ + S_- \sigma_-$ and $\delta S =0$
results in $\delta_+ S_+ =0$ plus $\delta_- S_- =0$.
{\it If we want to modify this,
obtaining a new field theory}, we are forced to extend the variational
equation, such that both parts are connected.
In \cite{schuller3,schuller0} the proposal is

\beqa
\delta S ~~~\epsilon~~~ \bd{P}^0 ~~~,
\eeqa
with $\bd{P}^0 = \bd{P}^0_+ \cup \bd{P}^0_-$ and
$\bd{P}^0_\pm = \left\{ X| X= \lambda \sigma_\pm \right\}$, i.e.,
the right hand side has to be in the zero divisor branch, which
plays in the pseudo-complex extension the r\^{o}le of a zero (remember that
the norm of $\lambda\sigma_\pm$ is zero).

Assuming now a theory with fields $\Phi_r$, we have for a 4-dimensional
space (1+3)

\beqa
\delta S & = & \int \left[ \sum_r \frac{D{\cal L}}{D\Phi_r} \delta \Phi_r
+ \sum_r \frac{D{\cal L}}{D(D_\mu \Phi_r)} \delta (D_\mu \Phi_r) \right]
d^4 X \nonumber \\
& = & \int \left[ \sum_r \frac{D{\cal L}}{D\Phi_r}
- \sum_r D_\mu \left(\frac{D{\cal L}}{D(D_\mu \Phi_r)} \right)
\right] \delta \Phi_r d^4X \nonumber  \\
& & + \sum_r \int D_\mu\left( \frac{D{\cal L}}{D(D_\mu\Phi_r)}
\delta \Phi_r \right) d^4X ~~~.
\eeqa

With $F^\mu (X) = \frac{D{\cal L}}{D(D_\mu\Phi_r)} \delta \Phi_r$,
the last term is surface integral

\beqa
\int (D_\mu F^\mu (X)) d^4X & = & \int (D_{+\mu} F_+^\mu (X)) d^4X_+ \sigma_+
\nonumber \\
& & 
+ \int (D_{-\mu} F_-^\mu (X)) d^4X_- \sigma_- \nonumber \\
& & \epsilon ~~~ \bd{P}^0 ~~~.
\eeqa
In standard field theory, this surface integral vanishes but here
we have to permit that the numerical result is a number in
the zero divisor branch $\bd{P}^0$.
This term can be added to the right hand side of the variational equation.
Without loss of generality, we assume that the element of $\bd{P}^0$
is of the form $\sum_r A_-^r \delta \Phi_{-,r}\sigma_-$, with some arbitrary
$A_-^r$.

From the variational equation we obtain

\beqa
D_{+,\mu} \left( \frac{D_+{\cal L}_+}{D_+(D_{+,\mu}\Phi_{+,r})} \right)
-\frac{D_+ {\cal L}_+}{D_+\Phi_{+,r}} & = & 0
\nonumber \\
D_{-,\mu} \left( \frac{D_-{\cal L}_-}{D_-(D_{-,\mu}\Phi_{-,r})} \right)
-\frac{D_- {\cal L}_-}{D_-\Phi_{-,r}} -A_-^r\sigma_- & = & 0 ~~~.
\eeqa
Or, in a compact notation,

\beqa
D_{\mu} \left( \frac{D{\cal L}}{D(D_{\mu}\Phi_{r})} \right)
-\frac{D {\cal L}}{D\Phi_{r}} & \epsilon & \bd{P}^0_- ~~~.
\label{eq-motion}
\eeqa
Strictly speaking, this is not an equation, though, we will continue
to denote it like that.
The same expression is obtained when we choose $\sum_rA_+^r\delta\Phi_{+,r}\sigma_+$
$\epsilon$ $\bd{P}^0_+$, being different from zero.
In order to obtain an equation of motion
of the type $\hat{A}=0$ one more step is involved as will be illustrated
next, on the level of a classical field theory.

\subsection{Scalar Fields}

The proposed Lagrange density is

\beqa
{\cal L} & = & \frac{1}{2} \left( D_\mu\Phi D^\mu\Phi - M^2 \Phi^2
\right) ~~~.
\label{l-phi}
\eeqa
The resulting equation of motion follows according to Eq. (\ref{eq-motion})
is

\beqa
(P^2 -M^2) \Phi & \epsilon & {\cal P}^0 ~~~.
\label{eqm0}
\eeqa
Multiplying by the pseudo-complex conjugate $(P^2-M^2)^*$ =
$(P_+^2 - M_+^2)\sigma_-$ + $(P_-^2 - M_-^2)\sigma_+$, we arrive at

\beqa
(P_+^2 -M_+^2)(P_-^2 -M_-^2)\Phi & = & 0 ~~~,
\label{eqm}
\eeqa
which can be seen as follows:
Without loss of generality we can assume the case
$(P^2 -M^2) \Phi ~ \epsilon ~ {\cal P}^0_-$. We have then

\beqa
(P^2 -M^2) \Phi & = & \left( (P_+^2 -M_+^2)\sigma_+ +
(P_-^2 -M_-^2)\sigma_- \right) \Phi \nonumber \\
& & ~\epsilon~ \bd{P}^0_- ~~~.
\label{p2m2}
\eeqa
This implies $(P_+^2 -M_+^2)\Phi = 0$, leading to
$(P_-^2 -M_-^2)\sigma_- \Phi$ = $(P_-^2 -M_-^2)\Phi_-\sigma_-$
= $\lambda \sigma_-$, with $\lambda$ having in general
some non-zero value. Alternatively,
$(P^2 -M^2)^*(P^2 -M^2)$ = $(P_+^2 -M_+^2)(P_-^2 -M_-^2)$
(use $\sigma_\pm^2=\sigma_\pm$, $\sigma_-\sigma_+ = 0$ and
$\sigma_+ + \sigma_- = 1$),
which is a pseudo-real hermitian operator whose eigenvalues are
real. It can only be satisfied when $\lambda = 0$.

{\it This is the connection we searched for: We
only obtain an equation of motion of the form $\hat{A}=0$, when both
components,
of the $\sigma_+$ and $\sigma_-$, are connected.}
The field equation is obtained, after having substituted $P_\pm^\mu$
by $p^\mu$ (see also comments at the end of section 3.1).

To obtain a solution for Eq. (\ref{eqm}), at least one of the factors, applied to
$\Phi$, has to vanish. Without loss of generality we choose the first one.
This defines which part of the pseudo-complex wave function we associate
with the standard physical particle.
After the above introduced extraction procedure, we obtain

\beqa
(p^2 -M_+^2)=0 ~\rightarrow~ E^2 = \bd{p}^2 + M_+^2 ~~~,
\eeqa
where we used $p^0=p_0=E$ and $p^k=-p_k$. It requires to interpret
the mass scale $M_+$ as the {\it physical mass} $m$ of the particle.

Eq. (\ref{eqm}), after setting $P^\mu_\pm = p^\mu$
and $X^\mu = x^\mu$, acquires the form

\beqa
(p^2 - M_+^2) \varphi (x) & = & 0 ~~~,
\eeqa
with the still pseudo-complex function, equal to
$\varphi_+(x) \sigma_+ + \varphi_-(x)\sigma_-$, and

\beqa
\varphi (x) & = & (p^2 - M_-^2) \Phi (x)
~~~.
\eeqa
This gives a relation of the field $\Phi (X)$ to
what we will call the {\it physical component}.

To obtain a value for the other mass scale $M_-$, 
we have to find a generalization of the propagator in this theory.
For that reason, let us consider the propagator
related to Eq. (\ref{eqm0}).
Its pseudo-complex Fourier component is

\beqa
\xi \frac{1}{P^2 - M^2} & = & \xi_+\frac{1}{P_+^2 - M_+^2} \sigma_+
+ \xi_-\frac{1}{P_-^2 - M_-^2} \sigma_-
\label{e55}
~~~,
\eeqa
where the factor $\xi = \xi_+\sigma_+ + \xi_-\sigma_-$
is in general pseudo-complex and has yet to be determined.
We used that $\frac{A}{B}$ = $\frac{A_+}{B_+}\sigma_+ +
\frac{A_-}{B_-}\sigma_-$.
Conversely, for Eq. (\ref{eqm}), setting
the pseudo-imaginary part of $P^\mu$ to zero, we expect the Fourier component

\beqa
\left( \frac{1}{p^2 - M_+^2} -
\frac{1}{p^2 - M_-^2} \right) ~~~.
\label{pauli-villars}
\eeqa

{\it In order to obtain a consistent result,
we have first to set in Eq.} (\ref{e55}) 
{\it $P^\mu_\pm$ to $p^\mu$ (selecting an inertial frame)
and taking the pseudo-real part.
In a second step, we have to
choose for $\xi$ the value} $2I$, {\it because} $I\sigma_\pm = \pm \sigma_\pm$.
Without the $I$,
the wrong relative sign appears and the factor of 2 is needed, to
get the correct normalization.
This result obtained will be resumed in section 4.4 within a formal
description of the extraction procedure.

The propagator describes the propagation of a particle characterized by
two mass scales. {\it In order to obtain the same result as SFT
{\it at low energy}, the $M_-$ has to be very large. Taking the analogy
to the Pauli-Villars regularization, $M_-$ should take the maximal
possible value, which is $l^{-1}$.}

The fact that a particle is described by two mass scales does
not imply two particles,
but rather the same particle with a dynamical mass, depending on the energy.

\subsection{Dirac Field}

The proposed Lagrange density for a Dirac particle is

\beqa
{\cal L} & = & \bar{\Psi} \left( \gamma_\mu P^\mu - M \right) \Psi ~~~.
\eeqa
The equation of motion is

\beqa
(\gamma_\mu P^\mu -M) \Psi & \epsilon & {\cal P}^0 ~~~.
\label{dirac1}
\eeqa
Multiplying by the pseudo-complex conjugate $(\gamma_\nu P^\nu-M)^*$ =
$(\gamma_\nu P_+^\nu-M_+)\sigma_-$ + $(\gamma_\nu P_-^\nu-M_-)\sigma_-$,
we arrive at

\beqa
(\gamma_\nu P_+^\nu -M_+)(\gamma_\mu P_-^\mu -M_-^2)\Psi & = & 0 ~~~.
\label{dirac-field}
\eeqa
Again, we have to project to the pseudo-real part of the momentum
in order to compare to the result of the SFT.
This leads us to assume that one
of the factors in Eq. (\ref{dirac-field}), applied to $\Psi$, has to be zero.
Without loss of generality we choose again the first one,
which describes a particle with physical mass $m=M_+$.

The Fourier component of the propagator,
corresponding to (\ref{dirac1}), is given by

\beqa
\xi\frac{1}{\gamma_\mu P^\mu - M} & = &
\xi_+\frac{1}{\gamma_\mu P_+^\mu - M_+} \sigma_+
+ \xi_-\frac{1}{\gamma_\mu P_-^\mu - M_-} \sigma_- ~~.
\nonumber \\
\eeqa
After projection, the expected form of the propagator, according
to Eq. (\ref{dirac-field}), is

\beqa
\left( \frac{1}{\gamma_\mu p^\mu - M_+} -
\frac{1}{\gamma_\mu p^\mu - M_-} \right) ~~~.
\label{green-dirac}
\eeqa
In order to be consistent with Eq. (\ref{dirac-field}),
requires to put $\xi =2I$, as in the scalar field case.
Like in the former example, the final operator describes the
propagation of a particle
with two mass scales.
In order to obtain at low energies the same result as in SFT,
again the $M_-$ has to be very large. We set $M_-$ equal
to the only mass scale left in the theory, which is $l^{-1}$.

{\it Note, that the theory is Pauli-Villars regularized. It is an automatic
consequence of the pseudo-complex description, i.e. the introduction
of a minimal invariant length.}

The dispersion relation for a Dirac particle is obtained
starting from Eq. (\ref{dirac-field}),
setting $P^{\mu}_2=0$,
multiplying it from the left with
$(\gamma_\nu p^\nu +M_-)$ $(\gamma_\mu p^\mu +M_+)$
and using the properties of the $\gamma^\mu$ matrices
($\frac{1}{2}\left( \gamma_\mu\gamma_\nu + \gamma_\nu\gamma_\mu \right)$ =
$g_{\mu\nu}$). The final result is
($M_+$ is renamed by $m$)

\beqa
(E^2 -\bd{p}^2 - m^2)= 0 ~~~.
\eeqa
As in the scalar case, we part from Eq. (\ref{dirac-field}), setting
$P^\mu_\pm = p^\mu$ and obtain

\beqa
(\gamma_\mu p^\mu - m)\psi (x) & = & 0 ~~~,
\eeqa
with

\beqa
\psi (x) & = & (\gamma_\mu p^\mu - M_-)\Psi (x) ~~~,
\eeqa
which gives the relation between $\Psi (X)$ to the physical projected piece.

\vskip 1cm
Let's summarize subsections 4.1 and 4.2: The procedure on how to deduce the
physical component is partially outlined, which states:
i) Set the pseudo-imaginary component of the linear momentum
to zero. This should also be the case for the pseudo-imaginary component
of the angular momentum, boost, etc.. ii) In order to get a propagator,
consistent with the field equation, we have to define the
Green's function by the equation

\beqa
\bd{O}G(X^\prime ,X) & = & (2\pi)^4 I \delta^{(4)}(X^\prime - X) ~~~,
\eeqa
where $\bd{O}$ = $(P^2 - M^2)$ for the scalar field case and
$(P{\sla} - M)$ for the pseudo-complex Dirac field (we use the
notation $P{\sla} = \gamma_\mu P^\mu$).
Only then
the correct sign appears, consistent with the field equation,
involving the two $\sigma_\pm$ components. The pseudo-complex
Fourier transform leads to the propagators discussed above.

The last piece of the extraction procedure, i.e.,
how the fields have to be mapped, will be discussed next.

\subsection{Extracting the physical component of a field}

To a pseudo-complex field $\Phi_r (X) = \Phi_{+,r}(X_+)\sigma_+ +
\Phi_{-,r}(X_-)\sigma_-$ a pseudo-complex mass is associated. We identified
the $M_+$ as the physical mass. When the motion of a {\it free}
particle is considered, the second component of the field, related to
the large mass $M_-=l^{-1}$, can not propagate on shell,
because for that at least the
energy $M_-$ is required. It contributes only to internal lines of
a Feynman diagram, where energy conservation is not required.
Therefore, the physical component of a {\it free} in- and out-going particle
is proportional to $\Phi_{+,r} (X)$,
where the pseudo-complex coordinate $X^\mu$
has to be subsequently substituted by $x^\mu$. This also holds for the linear momentum,
when the Fourier component of the field is considered.
In the case for the scalar field, the physical component is, therefore,
$\varphi_{+,r} (x)$ = ${\cal N} (p^2 - M_-^2)\Phi_{+,r}(x)$, with ${\cal N}$
a normalization factor. Taking into account
that $p^2\Phi_{+,r} (x)$ gives $M_+^2\Phi_{+,r} (x)$ = $m^2\Phi_{+,r} (x)$
(on-shell condition),
the factor $(m^2 - M_-^2)$ in front can be assimilated
into the normalization and we can use $\Phi_+ (x)$ as the projected
physical piece of the field. The same holds for the Dirac field.

For fields describing internal lines, similar to the propagators, the
physical component is given by the sum of the fields $\Phi_{\pm,r}$.
For example, when the dispersion of a charged particle at a Coulomb
field is considered, one has to take the sum of the $\sigma_\pm$
components of the pseudo-complex Coulomb field.

The discussion of this subsection leads to the last piece on how to
extract physical answers:
The construction of any composed object, like a S-matrix element
which is a function of the fields and propagators, is defined
by first constructing
the {\it projected pieces}, i.e., extract the pseudo-real part
of the propagators, take the $\sigma_+$ component of the in-
and out-going fields and then compose higher order expressions.
This is in analogy to Classical Electrodynamics, where the fields are
expressed as complex functions, but when non-linear expressions in the fields,
like the energy, are calculated, one has to decide which component to
take.

\subsection{Formal introduction to the extraction procedure}

Let us first propagate the pseudo-complex scalar field $\Phi (X)$
from the space-time point $X$ to $Y$
and then project the physical part of the wave function. It can be
written as

\beqa
\Phi (Y) & = & i \int d^4 X G(Y,X) \Phi (X) \nonumber \\
& \rightarrow & \varphi_+ (y)~=~{\cal N}(p^2 - M_-^2)\Phi_+(y)
\nonumber \\
& = & {\cal N} (p^2 - M_-^2)i\int d^4x G_+(y,x)\Phi_+(x)
~~~,
\eeqa
where ${\cal N}$ is a normalization factor and $Y$ has been set to $y$.
The Fourier transform of $G_+(y,x)$ is $\tilde{G}_+(p) =
\frac{1}{p^2 -M_+^2}$. Applying a Fourier transformation also
to the field $\Phi_+ (x)$, denoted as $\tilde{\Phi}_+(p)$, and
using the properties of the $\delta^{(4)} (p^\prime -p)$ function, we obtain

\beqa
\varphi_+(y) & = & i\int d^4p \frac{e^{ip\cdot y}}{p^2 - M_+^2}
\tilde{\varphi}_+(p) ~~~.
\label{varphi}
\eeqa

Now, we will first project and then propagate from $x$ to $y$. Because the
projected state is given by $\varphi_+(x)$ =
${\cal N} (p^2 - M_-^2) \Phi_+(x)$ it has to
be propagated by $g(y,x)$ with the Fourier transform
$\left[ 1/(p^2 - M_+^2) - 1/(p^2 - M_-^2) \right]$, as was
suggested above. Propagating this state gives

\beqa
& i\int d^4 g(y,x) \varphi_+(x) ~ = ~  & \nonumber \\
& \frac{1}{(2\pi)^2}\int d^4x \int d^4p_1 \int d^4p_2
\left( \frac{1}{p_1^2 - M_+^2} - \frac{1}{p_1^2 - M_-^2} \right) & \nonumber \\
& e^{ip_1 \cdot (y-x)} \varphi_+(p_2) e^{ip_2 \cdot x} ~= & \nonumber \\
& (M_+^2 - M_-^2)i\int d^4p \frac{e^{p \cdot y}}{(p^2-M_+^2)(p^2-M_-^2)}
\tilde{\varphi}_+(p) ~~~. &
\eeqa
exploiting the on-shell condition for a free particle
$p^2\varphi_+(p) = M_+^2\varphi_+(p)$, leads to the same
result as in Eq. (\ref{varphi}). Note, that we used the propagator
$g(y,x)$ for $\varphi_+ (x)$, while for $\Phi (X)$ it is $G(Y,X)$.
Using as the physical part of the wave function the $\varphi_+(x)$,
requires the physical propagator $g(y,x)$. Thus, a consistent formulation
is obtained.

\subsection{Conserved Quantities}

As in the SFT, the Noether theorem can be applied.
The procedure is completely analogous, except for appearance
of the zero divisor.

As a particular example, let us discuss the translation in space-time,
i.e., 

\beqa
X_\mu^\prime = X_\mu +\delta b_\mu ~~~,
\eeqa
where $\delta b_\mu$ is a constant four-vector. The variation of the
Lagrange density has to be proportional at most to a divergence plus
a term which is in the zero divisor branch:

\beqa
\delta {\cal L} & = & {\cal L}^\prime - {\cal L} ~=~ \delta b_\mu
D^\mu {\cal L} + \xi ~~~,
\eeqa
with $\xi ~\epsilon~ \bd{P}^0$.

Proceeding parallel to the SFT, using the equation
of motion (\ref{eq-motion}),
leads to \cite{greiner2}

\beqa
D^\mu \Theta_{\mu\nu} & \epsilon & \bd{P}^0
~~~,
\eeqa
with
\beqa
\Theta_{\mu\nu} & = & -g_{\mu\nu} {\cal L} +
\sum_r \frac{D{\cal L}}{D(D^\mu \Phi_r)} D_\nu \Phi_r
~~~,
\eeqa
which is the pseudo-complex energy momentum tensor. The $\Phi_r$ is
some field with an index $r$ and $g_{\mu\nu}$ is the metric involved.
Let us suppose that $\xi~\epsilon~ \bd{P}^0_-$.
When we look at the $\sigma_+$ component, we get an equation
$D_+^\mu \Theta_{+,\mu\nu} = 0$, which gives
the usual conservation laws, setting $P^\mu_\pm = p^\mu$.
Considering both components, the equation reads

\beqa
D^\mu \Theta_{\mu\nu} & \epsilon & \bd{P}^0 ~~~.
\eeqa

For the case of a scalar field,
this expression leads to the conserved linear momentum

\beqa
P^k & = & - \int d^3X \Pi (X) D^k \Phi (X) ~~~,
\eeqa
with $\Pi (X) = D^0\Phi (X)$, and the Hamilton density

\beqa
{\cal H} & = & \frac{1}{2} \left( \Pi^2 + (D^k \Phi )^2 + M^2 \right) ~~~.
\eeqa

Similar steps have to be taken for the Dirac and the electro-magnetic
fields and also when other symmetry operations, like rotations and
phase changes, are considered.

The symmetry properties of the fields are
similar to the SFT. Therefore, we will not elaborate
on them further.

\section{Gauge Symmetry and Gauge Bosons}

Let us consider the case of a Dirac particle, coupled to a photon
field, i.e., Pseudo-Complex Quantum Electrodynamic Field Theory (PSQED).
The proposed Lagrange density is \cite{schuller3}

\beqa
{\cal L} & = & \bar{\Psi} \left( \gamma^\mu (P_\mu - igA_{\mu})
- M \right) \Psi \nonumber \\
&& -\frac{1}{4}F^{\mu\nu}F_{\mu\nu} +\frac{1}{2}N^2\sigma_- A_\mu A^\mu
~~~,
\label{lagrange}
\eeqa
with $F_{\mu\nu} = D_\mu A_\nu - D_\nu A_\mu$, $D_{\mu}$ the pseudo-complex
derivative and $M$ being the pseudo-complex mass of the Dirac particle.
The photon has a pseudo-complex mass term given by $N^2\sigma_-$, i.e.,
the physical mass $N_+$ is zero.
Due to the appearance of a mass term,
one might worry about gauge invariance.
However, gauge invariance is still preserved in the pseudo-complex
formulation:

The fields transform as

\beqa
\Psi & \rightarrow & exp(i\alpha (x))\Psi \nonumber \\
A_{\mu} & \rightarrow & A_{\mu} +\frac{1}{g} \left( D_{\mu} \alpha (x)
\right) ~~~,
\eeqa
where $\alpha (x)$ is the gauge angle, depending on the position
in space-time. This gauge angle {\it can be chosen the same in both
parts, pseudo-real and pseudo-imaginary}. I.e., $\alpha (x)$ =
$\eta (x) (1 + I)$ = $(\eta (x) /2 )\sigma_+$. This is justified
because at the {\it same space-time} point an
observer can define the same gauge angle without violating
the principle of relativity. Therefore,
$\alpha (x)$ {\it gives zero when applied to the mass term} of the photon in
the Lagrange density. No contradiction to the principle
of gauge-symmetry exists!

The formulation can be extended to higher gauge symmetries, not
only $U(1)$, as just discussed.  For the case of an $SU(n)$ symmetry,
the Lagrange density of a Dirac field coupled to a gauge boson field
is given by ($a,b,c=1,2,...,n$)

\beqa
{\cal L} & = & \bar{\Psi} \left( \gamma^\mu (P_\mu - ig\lambda_a A^a_{\mu})
- M \right) \Psi \nonumber \\
&& -\frac{1}{4}F_a^{\mu\nu}F^a_{\mu\nu}
+\frac{1}{2}N^2\sigma_- A^a_\mu A_a^\mu
~~~,
\label{lagrange-un}
\eeqa
where $\lambda_a$ are the generators of the $SU(n)$ gauge group.
This Lagrange density was proposed in \cite{schuller3}, however,
without any further calculations.

Under gauge transformations they change to

\beqa
\Psi & \rightarrow & exp(i\alpha^a (x)\lambda_a)\Psi \nonumber \\
A^a_{\mu}\lambda_a & \rightarrow &
A^a_{\mu}\lambda_a +\frac{1}{g} \left( D_{\mu} \alpha^a (x)\lambda_a \right)
+i\alpha^a(x) A^b_\mu f_{ab}^c \lambda_c
~~~,
\eeqa
with $f_{ab}^b$ as the structure constants of the $su(n)$ algebra
(algebras are denoted by lower case letters).

\section{Quantization}

The quantization procedure will be illustrated first for the case
of a pseudo-scalar field. It is followed by the
Dirac field and finally by the electro-magnetic field.

\subsection{Scalar Field}

Note, that a pseudo-scalar field is
not a scalar with respect to the pseudo-complex numbers, because it
has non-vanishing components in $\sigma_\pm$. We refer rather to the scalar
nature with respect to the usual complex numbers.

In the first step of the quantization procedure we construct a
possible complete basis with respect to which we expand the pseudo-scalar
field $\Phi(X)$. Solutions to the above field equations
(\ref{eqm0}) and (\ref{eqm}) are plane waves

\beqa
f_P(X) & = & \frac{1}{(2\pi )^{\frac{3}{2}}\sqrt{2\omega_P}}e^{-iP \cdot X}
~~~,
\label{plane-w}
\eeqa
where $X$, $f_P(X)$, $P$ and $\omega_P$ are all pseudo-complex quantities.
The $\omega_P$ reflects the dispersion relation and is given by

\beqa
\omega_P & = & \sqrt{\bd{P}^2+M^2} ~=~
\sqrt{\bd{P}_+^2+M_+^2} ~\sigma_+ \nonumber \\
&& + \sqrt{\bd{P}_-^2+M_-^2} ~\sigma_-  ~~~.
\eeqa
The factor in Eq. (\ref{plane-w}) normalizes the components in each sector.

Since $f_P(X)$ is a solution to the field equation and
completeness can be shown in the same way as for pseudo-real
plane waves, the next step is to expand the field $\Phi (X)$ in
terms of these pseudo-complex plane waves:

\beqa
\Phi (X) & = & \int d^3P
\left[ \bd{a}(\bd{p}) f_P(X) + \bd{a}^\dagger (\bd{p}) \bar{f}_P(X) \right] ~~~.
\label{field-phi}
\eeqa
For the moment, the integration is taken along a straight line
(see Appendix A) in the pseudo-complex momentum space. Later, we will
restrict to the pseudo-real axis only. However, the more general form
has implications, discussed next. The physical consequences are not
completely clear
yet. Further investigation is needed.

One can deduce the commutation properties of the operators
$\bd{a}$ and $\bd{a}^\dagger$. This requires to assume a particular
commutation relation for equal times of the field with its canonical momentum, which
is defined as

\beqa
\Pi(X) & = & \frac{D{\cal L}}{D(D^0\Phi)} =D_0\Phi ~~~.
\eeqa

A general proposal for the commutation relation is

\beqa
\left[ \Phi (\bd{X},X_0), \Pi(\bd{Y},X_0) \right] & = &
iI^n\delta^{(3)} (\bd{X}-\bd{Y}) ~~~,
\label{phi-pi}
\eeqa
with $\delta(\bd{X}-\bd{Y})$ =
$\delta(\bd{X}_+-\bd{Y}_+)\sigma_+ + \delta(\bd{X}_--\bd{Y}_-) \sigma_-$,
as introduced in the Appendix A. The natural number $n$ has yet to be
specified.

In Appendix B the inversion of the above relations is given,
yielding the operators $a$ and $a^\dagger$ and their commutation relations:

\beqa
\left[ \bd{a}(\bd{P}),\bd{a}^\dagger (\bd{P}^\prime ) \right] & = &
I^{n+\xi_X} \delta^{(3)}(\bd{P} - \bd{P}^\prime) ~~~,
\label{a-com-a}
\eeqa
with $\xi_x$ related to the type of path chosen in integrations.
Conversely, let us start from the last equation, assuming the given commutation
relation, and deduce the commutation relation of the field with its
conjugate momentum, which should give back Eq. (\ref{phi-pi}).
As shown in Appendix B, this requires to set
$\xi_p = \xi_x$, i.e. if the integration in $X$ is in one sector, it has to
be in the equivalent one in $P$.

Let us suppose that $\xi_x=0$ (pure straight "space-like"
paths) and $n=0$ or $n=1$.
In the first case ($n=0$) we obtain the usual commutation relations,
{\it which we will adopt from now on}. As we
will show in the next subsection, this implies a particular definition of the
propagator in terms of the fields.

Relating the results to SFT, implies setting $P_\pm^\mu$ to $p^\mu$,
which gives for the component $\Phi_+ (x)$, now with $X^\mu$
$\rightarrow$ $x^\mu$, a plane wave proportional to $exp(ip_\mu x^\mu )$.
Therefore, an in- and out-going wave is described as before.

For completeness, we discuss the case with $n=1$: 
The commutation relation of the creation and annihilation operators
reduces to

\beqa
\left[ \bd{a}(\bd{P}),\bd{a}^\dagger (\bd{P}^\prime ) \right] & = &
I \delta^{(3)}(\bd{P} - \bd{P}^\prime) ~~~,
\label{comm-phi}
\eeqa
with all other commutators equal to zero.
Separating the commutator into the $\sigma_+$ and $\sigma_-$ part, where the
first is related to the low energy mass, also projecting
to real momenta, yields

\beqa
\left[ \bd{a}_+(\bd{p}),\bd{a}_+^\dagger (\bd{p}^\prime \right] & = &
+ \delta^{(3)}(\bd{p} - \bd{p}^\prime) \nonumber \\
\left[ \bd{a}_-(\bd{p}),\bd{a}_-^\dagger (\bd{p}^\prime \right] & = &
- \delta^{(3)}(\bd{p} - \bd{p}^\prime) ~~~.
\eeqa
The commutation relations for the $\sigma_-$ component have the
{\it opposite sign}, i.e., the part described by $\bd{a}_-$
seems to refer to a particle with unusual commutation relations.

Such structure is not new. In a formulation within Krein-spaces (see,
for example \cite{stumpf}), finite field theories
are obtained, suggesting a possible relation to our description.
The field equations look very similar to those discussed
in this contribution. When the particle with mass $M_+$ is considered
as the physical one, the commutation relations are the usual ones. However,
particles corresponding to the mass, $M_-$ in our case, obey
commutation relations with an opposite sign. These particles, within the
Krein-space formulation, never appear as free particles but rather only
in internal lines of the Feynman diagrams.
{\it Thus, choosing another path of integration in the pseudo-complex
space of $X^\mu$ and $P^\mu$ leads possibly to a different kind of theory}.
Mathematically, these theories are related and it would be
interesting to elaborate on it further.

\subsubsection{Propagator}

In section 4.1. the concept of a propagator for the scalar field was extended.
Using the standard commutation relations of the fields, the
creation and annihilation operators ($n=0$), the
following definition of a propagator, consistent with our former
discussion, can be given, namely

\beqa
I \langle 0 | \Phi (X) \Phi (Y) | 0 \rangle ~~~,
\label{green1}
\eeqa
assuming now that the field and their arguments are, in general,
pseudo-complex. We could have used the second choice of commutation
relations ($n=1$) as indicated in the previous subsection. Then, there would
be no factor $I$, implying unusual (anti-)commutation relations
and a different field theory.
{\it However, we prefer the standard commutation relations}, because they
allow the usual interpretation of the particles as bosons. The opposite
requires the introduction of particles with unusual properties,
as discussed above, but not observed.

Substituting the fields of Eq. (\ref{field-phi})
and using the commutator relations of the pseudo-complex
creation and annihilation operators (\ref{a-com-a}) gives

\beqa
I\int \frac{d^3P}{(2\pi )^{\frac{3}{2}}2\omega_P}
e^{-iP\cdot (X-Y)} ~~~.
\eeqa
For equal times ($Y_0 = X_0$) and $P_\pm = p$ we arrive at

\beqa
& I \langle 0 | \Phi (X) \Phi (Y) | 0 \rangle = & \nonumber \\
& I\left\{
\frac{1}{(2\pi )^3} \int \frac{d^3P}{2\omega_{+,p}}
e^{-i\bd{p}\cdot (\bd{x}-\bd{y})} \sigma_+ \right. & \nonumber \\
& \left. + \frac{1}{(2\pi )^3} \int \frac{d^3P}{2\omega_{-,p}}
e^{-i\bd{p}\cdot (\bd{x}-\bd{y})} \sigma_- 
\right\} & ~~~.
\eeqa
This must still be projected to the pseudo-real part, which is
the sum of the factor of $\sigma_+$ and $\sigma_-$.
Due to the $I$ as a factor and $I\sigma_\pm = \pm\sigma_\pm$, the
sign in the second term changes and we obtain the propagator
of Eq. (\ref{pauli-villars}). This is possible having chosen the
quantization, with $n=0$, as given above.

As can be seen, the description is consistent, using the proposed form
of the propagator in Eq. (\ref{green1}). The advantage lies in the standard
use of the commutation relations of the fields, the creation and annihilation
operators and its interpretation as bosons.

\subsection{Dirac Field}

The quantization of the Dirac field has the usual form \cite{greiner},
using now $E_P=\omega_P$,

\beqa
\Psi (X) & = & \sum_{\pm s} \int \frac{d^3P}{(2\pi )^{\frac{3}{2}}}
\sqrt{\frac{M}{E_P}}
\left[ \bd{b}(P,s) u(P,s) e^{-iP\cdot X} \right. \nonumber \\
& & \left. + \bd{d}^\dagger (P,s) v(P,s) e^{iP\cdot X} \right]
\nonumber \\
\Psi^\dagger (X) & = & \sum_{\pm s} \int \frac{d^3P}{(2\pi )^{\frac{3}{2}}}
\sqrt{\frac{M}{E_P}}
\left[ \bd{b}^\dagger (P,s) \bar{u}(P,s) e^{iP\cdot X} \right. \nonumber \\
& & \left. + \bd{d} (P,s) \bar{v}(P,s) e^{iP\cdot X} \right]
~~~,
\eeqa
with the exception that all functions and operators are pseudo-complex.
The bar over a function indicates the normal complex conjugation.
The $s$ indicates the two possible spin directions and
$E_P = \sqrt{\bd{P}^2 + M^2}$.

The anti-commutator relations at equal time are set to

\beqa
\left\{ \Psi (\bd{X},X_0) , \Psi^\dagger (\bd{Y},X_0) \right\} & = &
I^n\delta^{(3)} \left( \bd{X} - \bd{Y} \right) ~~~.
\eeqa
All other anti-commutators are zero. There are the
two possibilities, $n=0$ or $n=1$. The case $n=0$ leads to standard
anti-commutation relations, while $n=1$ leads to the commutation
relations as discussed in \cite{stumpf}. {\it We choose,
as in the boson case,
the standard anti-commutation relations}.

The result is

\beqa
\left\{ \bd{b} (P,s) , \bd{b}^\dagger (P^\prime ,s^\prime ) \right\}
& = & \delta_{ss^\prime} \delta^{(3)}\left(\bd{P}^\prime -  \bd{P}\right)
\nonumber \\
\left\{ \bd{d} (P,s) , \bd{d}^\dagger (P^\prime ,s^\prime ) \right\}
& = & \delta_{ss^\prime} \delta^{(3)}\left(\bd{P}^\prime -  \bd{P}\right)
~~~,
\eeqa
and all other anti-commutation relations equal to zero.

Again, the propagator of the form (\ref{green-dirac}) is only
obtained when in terms of the fields it is defined as

\beqa
I\langle 0 | \Psi (X^\prime ) \Psi (X) | 0 \rangle ~~~.
\eeqa

Also here, the in- and out-going states are obtained by first
mapping $P_\pm^\mu$
$\rightarrow$ $p^\mu$ and $X_\pm^\mu$ $\rightarrow$ $x^\mu$.
The field is then a simple in- and out-going plane wave, multiplied with
the corresponding Dirac-spinor.

\subsection{The Electro-Magnetic Field}

The procedure is completely analogous to the one outlined in the two
last cases. The quantized electro-magnetic field is \cite{greiner}

\beqa
\bd{A}(X) & = &
\int \frac{d^3P}{(2\pi)^3 2\omega_P} \sum_{\lambda=1,2} \bd{e}(P,\lambda )
\left[ \bd{a}(P,\lambda ) e^{-iP\cdot X} \right. \nonumber \\
& & \left. + \bd{a}^\dagger (P,\lambda ) e^{iP\cdot X} \right] ~~~. 
\eeqa
The interpretation is analogous to the usual field theory, with the
exception that the fields and variables are now pseudo-complex.
The $\lambda$ indicates the two polarization directions
and $\bd{e}(P,\lambda )$ are the unit vectors of the
polarization $\lambda$.

As in the scalar and Dirac field cases, the in- and out-going waves
are proportional to $\bd{e}^\mu exp(\mp i p_\mu x^\mu)$.

The quantization rule for equal pseudo-complex time is

\beqa
\left[ \Pi_i (\bd{X},X_0) , A^j (\bd{Y},X_0) \right] & = &
i\delta^{(tr)}_{ij} \left( \bd{X} - \bd{Y} \right) \nonumber \\
& = & i\left[ \delta^{(tr)}_{ij} \left( \bd{X}_+ - \bd{Y}_+ \right) \sigma_+
\right. \nonumber \\
&& \left.
+ \delta^{(tr)}_{ij} \left( \bd{X}_- - \bd{Y}_- \right) \sigma_- \right]
~~~,
\nonumber \\
\eeqa
with the {\it transversal} delta function on the right hand side of
the equation.

The pseudo-scalar mass of this field has a zero $\sigma_+$ component,
which is related to the zero physical rest mass at low energy.
The $\sigma_-$ component has to be large, as argued also in the
case of a bosonic and fermionic field. Again, the only
mass scale left, equal to $l^{-1}$, is taken for the $\sigma_-$
component, denoted by $N$.
It is reflected in the dispersion relations $\omega_P$ =
$\omega_{+,P} \sigma_+ + \omega_{-,P} \sigma_-$, with
$\omega_{+,P}=P_+$ and $\omega_{-,P}=\sqrt{P_-^2 + N^2}$. This choice leads,
with the additional $I$ in the definition of the pseudo-complex propagator,
to

\beqa
\frac{1}{P_+^2} \sigma_+ - \frac{1}{P_-^2 - N^2} \sigma_- ~~~.
\eeqa
Setting $P_\pm^\mu = p^\mu$ and extracting the
pseudo-real part, leads to

\beqa
\frac{1}{p^2} - \frac{1}{p^2 - N^2} ~~~,
\label{ef-mass}
\eeqa
i.e., to the desired result of the propagator.

A consequence of (\ref{ef-mass}) is an effective mass
of the photon as a function in energy. We set
the propagator in (\ref{ef-mass}) equal to $1/(p^2-m(\omega )^2)$,
with $m(\omega )$ being a {\it effective rest mass} at a fixed energy
$\omega$. Solving for $m(\omega )$ yields $p^2/N$. Setting $p^2$
equal to $\omega^2$, gives

\beqa
m(\omega ) & = & \frac{\omega^2}{N} ~=~ l\omega^2 ~~~.
\label{mass}
\eeqa

At energies in the GeV range, the $m(\omega)$ is of the order of
$10^{-20}$~GeV, far to low to be measured. Thus, the photon appears to
have no mass. At energies of $10^{11}$~GeV, the scale of the
GZK limit, this mass rises to
about 500~GeV. It sounds large, but it has to be compared with the
the energy scale, giving a ratio of about $5*10^{-9}$.

This has a measurable effect on the dispersion relation.
The energy of the photon is given by

\beqa
\omega^2 & = k^2 + m(\omega)^2 ~~~,
\eeqa
where we used ($\hbar = 1$) $E=\omega$ and $p=k$.
Solving for $\omega$, using
Eq. (\ref{mass}) leads in lowest order in $l$ to ($N=l^{-1}$)

\beqa
\frac{\omega}{k} & = & 1- \frac{1}{2}(lk)^2 ~~~,
\eeqa
which shows the deviation from the light velocity. For energies
at the GZK scale ($\omega = 10^{11}$~GeV) and using $l=5*10^{-20}$~GeV$^{-1}$,
the second term on the right hand side acquires the value of
$2.5*10^{-18}$, still quite low. For energies of the order of
$50$~TeV = $50000$~GeV the effect is even smaller, about $10^{-34}$. In
\cite{stecker} upper limits on the correction to the speed of light
were deduced for energies in the TeV range, using experimental
observations. The stringenst limit, obtained for the case of Compton
scattering of photons in the TeV range, is $< 10^{-16}$. 

For a free propagating photon, the effect of the
mass $N$ via the vacuum polarization can be assimilated in
a renormalization of the charge, as shown in \cite{greiner}.
It results in a modification, due to
the mass scale $l^{-1}$, 
to the dependence on the energy scale in the running
coupling constant. Thus,
renormalization is still necessary, although it is not
related to erase infinities any more.

There is an interesting interpretation of the zero component
of the vector potential. Using the propagator and searching for
the Fourier transform, gives for the $\sigma_+$ part a simple
Coulomb potential $-\frac{1}{r_+}$,
while the $\sigma_-$ part describes a propagating particle
with mass N, i.e., it results in a Yukawa potential $e^{-N r_-}{r_-}$.
Projecting to the pseudo-real part, with $r_\pm = r$, gives

\beqa
A_0 & \sim & -\frac{1}{r} \left( 1 - e^{-N r} \right) ~~~.
\label{pseudo-coulomb}
\eeqa
For large $r$ it is essentially the Coulomb potential. However,
for $r$ of the order of $\frac{1}{N}\sim l$ a deviation appears.
For $r \rightarrow 0$ we get $A_0 \rightarrow \sim -N$,
which is a large number.

\section{Calculation of some Cross Sections}

We will determine two different cross sections: a) the dispersion
of a charged particle in an external Coulomb field and b) the
Compton scattering. The steps are analogous to the ones described in
\cite{greiner}. The two cross sections chosen, differ
to lowest order in the internal lines, which is a photon in the
first case and a fermion in the latter.

We use the proposed projection method on how to
extract numerical results. It requires the
construction of the building
blocks of the $S$-matrix elements, i.e., the $\sigma_+$ component of
the fields and the pseudo-real part of the
propagators, and then compose the $S$-matrix element.
When a filed appears in internal lines, it is treated similar
to the propagators, i.e., the sum of the $\sigma_\pm$ components have
to be taken.
The cross section is proportional to the square of the $S$-matrix
element.

We take into account only electro-magnetic interactions.
The united electro-weak field theory should be used,
because the interesting deviations will probably
occur at very large energies. This would, however,
explode the scope of the present contribution. To get more realistic cross
sections at ultra-high energies, we refer to a later publication.

\subsection{Scattering of a charged Particle at a Coulomb Potential}

We proceed in a completely analogous way
as in Ref. \cite{greiner}. The transition matrix element
is given by

\beqa
S_{fi} & = & -ie \int d^3X \bar{\Psi}_f(X) \gamma^\mu A_\mu (X) \Psi_i (X)
~~~,
\label{sif}
\eeqa
where the indices $i$ and $f$ refer to the initial and final state
respectively. The fields in Eq. (\ref{sif}) are substituted
according to the rules formerly established.

The in- and out-coming field are given by

\beqa
\Psi_i (x) & = & \sqrt{\frac{m}{E_iV}} u(p_i,s_i) e^{-ip_i \cdot x}
\nonumber \\
\bar{\Psi}_f (x) & = & \sqrt{\frac{m}{E_fV}}
\bar{u}(p_f,s_f) e^{-ip_f \cdot x} ~~~,
\eeqa
with $E_{i/f} = \sqrt{p^2_{i/f} + m^2}$.
The Coulomb field describes the mediating photons and one has to
take the pseudo-real component of

\beqa
A_0(X) & = & 2\left[ -\frac{Ze}{4\pi |X_+|} \sigma_+
+ \frac{Ze}{4\pi |X_-|}e^{-N|X_-|} \sigma_- \right]
\label{a0}
~~~.
\eeqa
Determining the partial cross section involves integration over the
coordinates, which we assume to be along the pseudo-real axis,
i.e., $|X|=r$, and $P_\pm = p$.

Taking the pseudo-real part of (\ref{a0}),
leads to the transition matrix element

\beqa
S_{fi} & = &
iZe^2\frac{1}{V}\sqrt{\frac{m^2}{E_fE_i}}
\frac{\bar{u}(p_f,s_f)\gamma^0 u(p_i,s_i)}{|\bd{q}|^2} 2\pi \delta (E_f-E_i)
\nonumber \\
& & - iZe^2\frac{1}{V}\sqrt{\frac{m^2}{E_fE_i}}
\frac{\bar{u}(p_f,s_f)\gamma^0 u(p_i,s_i)}{|\bd{q}|^2+N^2} 2\pi \delta (E_f-E_i)
~~~,
\nonumber \\
\eeqa
with $\bd{q}=\bd{p}_f-\bd{p}_i$.
The mass $N$, is the $\sigma_-$ component of the photon's
pseudo-complex mass.

One finally arrives at the expression

\beqa
\frac{d\sigma}{d\Omega}& = &
4Z^2\alpha^2m^2 \left( \frac{1}{|q|^2} - \frac{1}{|q|^2+N^2} \right)^2
\nonumber \\
& & |\bar{u}(p_f,s_f)\gamma^0 u(p_i,s_i)|^2
~~~.
\eeqa

Using the mean value of the cross section for different spin
orientations \cite{greiner} and the kinematic relations $E_i=E_f=E$
(elastic scattering) $\left( \bd{p}_i \cdot \bd{p}_f \right)$ =
$E^2 -p^2 cos\theta$, we arrive at

\beqa
\frac{d\bar{\sigma}}{d\Omega} & = & \frac{Z^2\alpha^2}{4p^2\beta^2
(sin\frac{\theta}{2})^4}\left( 1-\beta^2sin^2\frac{\theta}{2} \right)
\nonumber \\
&&\left[ 1-\frac{p^2sin^2\frac{\theta}{2}}{N^2+4p^2sin^2\frac{\theta}{2}}
\right]^2 ~~~.
\eeqa
The bar over the $\sigma$ indicates the summation over the spin directions
of the in- and out-coming particle.
The factor in front of "$\left[ ... \right]^2$" is the Mott formula
for the scattering of an electron at a Coulomb potential of a nucleus.

Considering that $N=\frac{1}{l}$, we get
to lowest order in $l$

\beqa
\frac{d\bar{\sigma}}{d\Omega} & \approx &
\frac{d\bar{\sigma}}{d\Omega}\left|_{{\rm Mott}}\right.
\left[ 1- 8l^2 p^2 sin^2\frac{\theta}{2} \right] ~~~.
\eeqa

The largest correction is at back scattering ($\theta = \pi$).
However, even for linear momenta near the GKZ cutoff
($p\approx 10^{11}$~GeV), the corrections are of the order of $10^{-16}$ 
($l\approx 5~10^{-20}$GeV$^{-1}$, corresponding to the Planck length),
beyond any hope to be measured in near future.
At momenta in the TeV range, the situation is even more hopeless.
The corrections would be of the order of $10^{-31} - 10^{-32}$.

\subsection{Compton Scattering}

The calculation of the cross section proceeds in the same way as
explained in \cite{greiner}. Traces of $\gamma$-matrices appear
which are of the form (we use the Dirac notation
$A{\sla} = \gamma_\mu A^\mu$)

\beqa
\bd{B}_{\sigma_1\sigma_2} & = &
{\rm Tr}\left[ \frac{p{\sla}_f + m}{2m} \Gamma_{\sigma_1}
\frac{p{\sla}_i + m}{2m} \bar{\Gamma}_{\sigma_2} \right] ~~~,
\eeqa
with $\sigma_k=\pm$ and

\beqa
\Gamma_\pm & = &
\frac{\epsilon{\sla}^\prime
\left(
p{\sla}_i + k{\sla}+M_\pm \right) \epsilon{\sla}}
{2p_i\cdot k + (m^2 -M_\pm^2)}
\nonumber \\
&& + \frac{\epsilon{\sla}
\left(
p{\sla}_i - k{\sla}+M_\pm\right)\epsilon{\sla}^\prime}
{-2p_i\cdot k + (m^2 -M_\pm^2)} 
\eeqa
and $\bar{\Gamma}_\sigma =\gamma^0 \Gamma_\sigma \gamma^0$.
We use $M_+=m$ and $M_- = \frac{1}{l}$. For the plus sign
we get the usual expression. The two possibilities of $\Gamma_\pm$
appear because to the propagator $1/(p{\sla} - m)$ of the SFT
one has to add the second term $-1/(p{\sla} - M_-)$.

When the minus index is used, we can exploit the large value of
$M_->>p_i,p_f,k$ and approximate $\Gamma_-$ through

\beqa
\Gamma_- & \approx &- \frac{2(\epsilon\cdot\epsilon^\prime )}{M_-}
~~~.
\eeqa

We arrive finally at the expression for the total cross section,
having made the usual relations between $p_i$ and $p_f$ \cite{greiner}
and evaluate the cross section in the laboratory frame, with
$p_i=(m,0,0,0)$. We obtain

\beqa
\frac{d{\bar \sigma}}{d\Omega_k^\prime} (\lambda^\prime ,\lambda )
& \approx &
\frac{1}{4m^2}\alpha^2 \frac{\omega^{\prime 2}}{\omega^2}
\left\{
\frac{\omega^\prime}{\omega} + \frac{\omega}{\omega^\prime}
+4(\epsilon\cdot\epsilon^\prime )^2 -2 \right\}
\nonumber \\
& & -4\frac{m}{M_-} \left\{
\frac{1}{m}(\epsilon\cdot\epsilon^\prime )(\epsilon\cdot k^\prime )
(\epsilon^\prime\cdot \epsilon ) \left( \frac{1}{\omega}-\frac{1}{\omega^\prime}
\right) \right. \nonumber \\
&& \left. + (\epsilon\cdot\epsilon^\prime )^2 \left(
\frac{\omega^\prime}{\omega} + \frac{\omega}{\omega^\prime} + 2 \right)
\right\}
~~~.
\eeqa
Summing over the initial polarizations ($\lambda$, $\lambda^\prime$)
of the photons we arrive at

\beqa
& \frac{d{\bar \sigma}}{d\Omega_k^\prime}   \approx & \nonumber \\
& \frac{\alpha^2\omega^{\prime 2}}{2m^2\omega^2}
\left( \frac{\omega^\prime}{\omega} + \frac{\omega}{\omega^\prime}
-sin^2(\theta ) \right)  & \nonumber \\
& \left\{ 1- \frac{m}{M_-}
\frac{
\left[ \frac{(\omega^\prime - \omega )}{m}cos\theta sin^2\theta
+(1+cos^2\theta ) \left( 
\frac{\omega^\prime}{\omega} + \frac{\omega}{\omega^\prime}+2\right)\right]
}{
\left[ \frac{\omega^\prime}{\omega} + \frac{\omega}{\omega^\prime}
-sin^2(\theta ) \right]
} \right\} &
~~~.
\nonumber \\
\eeqa

As can be seen, the correction is proportional to $\frac{m}{M_-}=ml$.
The deviations are increased when heavy particles, like $W^\pm$ and $Z$
bosons are involved. Choosing $\theta$ in the forward or backward scattering
and using a particle of mass $\approx 100$~GeV, leads to a
correction of the order of (using also $\omega^\prime \approx \omega$)
$-100/(5*10^{20})$ = $2*10^{-19}$, which is still low, but easier to
measure than in the Coulomb scattering of a charged particle.

Obviously, an internal photon line gives as the smallest correction terms
proportional to $l^2$, while an internal electron line gives a correction
proportional to $l$.
This is due to the dependence of the propagator on $M_-$, which is
$\sim (1/M_-^2)$ for the photon and $\sim (1/M_-)$ for the fermion.
If one searches for detectable deviations one
should, therefore, choose processes which include dominantly electron
internal lines.

\subsection{Lamb Shift and magnetic Moment of the electron}

We also looked at possible changes in the Lamb shift and the magnetic
moment of the electron. After having applied the charge and mass
renormalization \cite{greiner}, the main corrections come from the internal
photon line and it is proportional to $q^2l^2$
= $q^2*25*10^{-40}$, with $q$ being the
momentum exchange. Because of the smallness of $l$ and $q$, the corrections
are far less than the current accuracy of about $10^{-11}$ in, e.g., the
anomalous magnetic moment of the electron.

Thus, the appearance of
a smallest length scale in the pseudo-complex field theory does not
result in measurable effects, considering standard high precision
experiments. The only hope to see a difference is the possible observation
of the GZK cutoff.

\section{Relation to Geometric Approaches}

It is illustrative to show a connection to geometrical
formulations in flat space-time, especially those
which are related to our approach.
It also will give a hint on how to extend the pseudo-complex
field theory such that it permits a shift of the GZK limit.
The language will be held simple in order to avoid unnecessary
complex notations.

Caianiello showed in (1981) \cite{caneloni} the existence of a maximal
acceleration, by combining the metric
in the {\it coordinate} and {\it momentum}
space.
This metric is very similar to the one in Eq. (\ref{pseudo-complex-metric})
below, for the pseudo-complex description.
He argued for this combination of position and momentum
in the same line element due to the uncertainty relation which treats
momentum and coordinate on an equal footing. This was already observed
by M. Born \cite{born1,born2}, called now the reciprocity theorem of Born.

To show more details,
let us define an {\it orbit} in the space-time manifold by

\beqa
X^\mu & = & x^\mu + lI u^\mu ~~~,
\label{xmu}
\eeqa
where $u^\mu = \frac{dx^\mu}{d\tau}$ is the four-velocity,
$l$ the invariant length scale and
$\tau$ the proper time. The $l$ appears for dimensional reasons.
It is a generalized notation also encountered and justified in section 3.1,
including, besides the position
of the observer, the information about his four velocity.
In Eq. (\ref{xmu}),
the {\it observer} is not only characterized by its position
$x^\mu$ in the Minkowski space, but also by its four-velocity, which defines
a 4-bein along the world line he realizes in the Minkowski space.
Eq. (\ref{xmu}) is a possibility to unite in one coordinate the position
$x^\mu$ of the observer with the {\it co-tangent}
space, given by the 4-bein, defined through $u^\mu = \frac{dx^\mu}{d\tau}$.
The geometrical implications are much more involved, related
to the fiber bundle description on which we will not elaborate here.

In a similar way the four velocity $u^\mu = \frac{dx^\mu}{d\tau}$
and the four momentum $p^\mu = \gamma m a^\mu$ are modified to

\beqa
U^\mu & = & u^\mu + lIa^\mu \nonumber \\
P^\mu & = & p^\mu + lIf^\mu ~~~,
\eeqa
with $a^\mu = \frac{du^\mu}{d\tau}$ as the four-acceleration and
$f^\mu$ as the four-force. $U^\mu$ is obtained through the derivation
of $X^\mu$ with respect to the eigen-time.

The scalar product with respect to the $dX^\mu$,
defines a new line element, given by
\cite{schuller2,schuller0}

\beqa
d\omega^2 & = & \eta \left(dX,dX\right) \nonumber \\
& = & dX^\mu dX_\mu = dx^\mu dx_\mu +l^2du^\mu du_\mu \nonumber \\
& & + lI (2dx^\mu du_\mu)  ~~~.
\label{pseudo-complex-metric}
\eeqa
The $d\omega$ is also considered as a generalized proper time.
Now, $x^\mu u_\mu$ = $ x^\mu \frac{dx^\mu}{d\tau}$ = 0. Therefore, it
can be rewritten as (with $u^\mu = \frac{dx^\mu}{d\tau}$,
$a^\mu = \frac{du^\mu}{d\tau}$ and $d\tau^2=dx^\mu dx_\mu$),

\beqa
d\omega^2 & = & d\tau^2 (1-l^2 g(u,a)) \nonumber \\
& & g(u,a)= -a^\mu a_\mu=-a_0^2+a_k^2 ~~~.
\label{omega-2}
\eeqa
Using $l^2 g(u,a)$ = $ \frac{a^2}{G_0^2}$, with $G_0=\frac{1}{l^2}$,
and requiring the positive definiteness of the metric ($d\omega^2>0$),
we arrive at the maximal acceleration $G_0$.

The new proper time $d\omega$ is related to the standard eigen-time
$d\tau$ via

\beqa
d\omega & = & \sqrt{1-l^2g^2}~d\tau ~~.
\eeqa
The factor in front of $d\tau$ reflects an additional $\gamma$ factor,
added to the one in special relativity.

In contributions based on a geometric description
\cite{schuller3,low1,beil1},
one usually defines the $d\omega^2$
as

\beqa
d\omega^2 & = & dx^\mu dx_\mu +l^2 du^\mu du_\mu 
\label{finsler-metric}
\eeqa
alone.
This metric is invariant under transformations of $O(2,6)$ (the measure
contains two time and 6 space components: $dx^0$, $du^0$ and
$dx^k$, $du^k$). Comparing this with the pseudo-complex metric, the
difference is in the term $2lI(dx^\mu du_\mu$) = 0. This might be
irrelevant. However, as stated in \cite{schuller4,schuller0}, its
omission leads
to a contradiction to the Tachibama-Okumara no-go theorem \cite{tach}.
It states that when the space-time manifold has an almost complex structure,
as the theories published in \cite{low1,beil1} have, a parallel transport
does not leave invariant this structure. In contrast, when the line element is
chosen as in (\ref{pseudo-complex-metric}), the space-time manifold has
an almost product structure and the Tachibama-Okumara theorem is satisfied.
However, in \cite{low1,beil1} the symplectic structure, which produces the
almost complex structure, is essential in order to maintain the commutation
relations between the coordinates and momenta. This indicates that there
are still important points not understood, which
have to be studied yet.

In \cite{low1,low2} the representation theory is discussed,
allowing only canonical, symplectic transformations, $Sp(4)$.
This restriction is necessary in order to maintain the
commutation relation of the coordinates with the momenta invariant.
The intersection is the
group $U(1,3)\simeq O(2,6) \cup Sp(4)$. Including translations, Low
arrives at what he denominates as the group $CR(1,3) \simeq U(1,3)\otimes_s H(1,3)$, where
$\otimes_s$ refers to the semi-direct product and $H(1,3)$ to the
Heisenberg group in 1+3-dimensions. For details, consult the
references \cite{low1,low2}.

Beil \cite{beil1,beil2,beil3,beil4} included the electromagnetic
gauge potential into the metric, showed the connection to
a maximal acceleration and claims a relation to Finslerian metrics.
The approach by Beil, however, is put on doubt \cite{goenner4} due to several
inconsistencies, like identifying the energy with the Lagrangian at
one point and mixing notations in \cite{beil1}.

There are several other geometrical approaches, where
the relation to ours is not so clear up to now:
Brandt \cite{brandt1,brandt2,brandt3},
developed a geometric formulation, including gauge
transformations. All gauge fields are included in the metric.

We also mention different geometrical approaches, based on
the pseudo-complexification of geometry.
To our knowledge, the first person, who proposed
this extension is A. Crumeyrolle. In \cite{crumeyrolle} pseudo-complex
numbers where introduced (called by him hyperbolic numbers) and
the coordinates $x^\mu$ of space-time were complexified hyperbolically.
A couple further contributions appeared \cite{crum2,crum3} and
other authors continued on this line
\cite{moffat1,kunstatter,moffat2,mann}. The theory presented has some
mathematical overlap to our formulation but the basic physical description and philosophy
are different.

As a last example of the geometric approach we mention
Refs. \cite{goenner1,goenner2,goenner3}. They introduce a preferred
velocity, thus, breaking rotational invariance explicitly. The Poincar\'e
group breaks down to a subgroup with 8 generators. Lorentz invariance
is explicitly broken and it is proven to be related to a Finslerian metric.
Three different solutions of possible metrics are discussed,
corresponding to a space-, time- and light-like velocity vector.
How this is related to the pseudo-complex description is not clear yet.
Only in the former mentioned approaches a relation is presented for
a flat space-time manifold.

Conclusion: {\it In flat space-time there is a correspondence to
some geometric approaches previously discussed in the literature.} They will
give a hint on how to extend the extraction to physically observable numbers.

\section{Extension of the Pseudo-Complex Field Theory}

The last section contains a hint on how possibly to extend the
pseudo-complex field theory. It is related to a modification
of the extraction procedure.
Up to now, the $P^\mu$ has two linear independent components,
namely the pseudo-real $P_1^\mu$ and the pseudo-imaginary
$P_2^\mu$ one. In the last section we saw, however, that
one can interpret the pseudo-imaginary component in a consistent way as
a force, acting on the particle along its world line. This can be
seen as a {\it projection} to a subspace
in the pseudo-complex space of $P^\mu$, with the constriction of
$P_2^\mu = lf^\mu$. Therefore, instead of setting the
pseudo-imaginary component to zero, it is {\it substituted} by $lf^\mu$.
This is equivalent to add to the Lagrange density an additional term,
reflecting the effect of the particle's acceleration
during the interaction. But it is
more: This interaction originates as a
part of the pseudo-complex linear momentum and,
thus, represents an extension of the minimal coupling scheme to the
pseudo-complex formulation:

\beqa
P_\mu & \rightarrow & p_\mu + lI f_\mu ~~~.
\eeqa

We can then proceed in the same way as done in \cite{hess1}, where the first
results of the PCFT, related to the shift of the GZK limit,
were presented. The equation of motion for a Dirac particle changes to

\beqa
\left( \gamma^\mu (p_\mu + lIf_\mu) - M \right) \Psi ~~~\epsilon ~~~
{\cal P}^0 ~~~,
\eeqa
with ${\cal P}^0$ = ${\cal P}_+^0 \cup {\cal P}_-^0$, is the set of
zero divisors. The $f_\mu$ may contain a dependence on the photon
field, but not necessarily.

Using $P_\pm^\mu = p^\mu \pm lf^\mu$,
multiplying by the pseudo-complex conjugate of the operator
gives $(P{\sla}_+ - M_+)(P{\sla}_- - M_-)\Psi = 0$, and subsequently
multiplying by $(\gamma_\mu \left[ p^\mu - lf_\mu \right] + M_-)$
$(\gamma_\mu \left[ p^\mu  + lf_\mu \right]+ M_+)$ and
using the properties of the $\gamma^\mu$ matrices,
we arrive at the equation

\beqa
\left( P_{+\mu} P^{\mu}_+ - M_+^2 \right)
\left( P_{-\mu} P^{\mu}_- - M_-^2 \right) \Psi & = & 0 ~~~.
\eeqa
Selecting the first factor, using
$P_{+\mu} P^{\mu}_+= E^2-p^2+l^2f_\mu f^\mu$
$+$ $l(p_\mu f^\mu + f_\mu p^\mu)$, we arrive at the
dispersion relation

\beqa
E^2 = p^2 + (lf)^2 +l(pf+fp)+M_+^2 ~~~,
\eeqa
with $f^2=-f_\mu f^\mu >0$.
and $pf=-p_\mu f^\mu$, $fp=-f_\mu p^\mu$.
When $f^\mu$ is a force, it is proportional to $\frac{dp^\mu}{d\tau}$
and, thus, $pf=fp=0$.

This leads to a modification of the threshold momentum, for the production
of pions in  a collision of a proton with a photon from the CMB,
\cite{hess1}

\beqa
p_{1,{\rm thr.}} & \approx & \frac{(\tilde{m}_2+\tilde{m}_3)^2-
\tilde{m}_1^2}{4\omega}
~ \approx ~ \frac{(m_2+m_3)^2-m_1^2}{4\omega}  \nonumber \\
& & +\frac{l^2}{4\omega} \left[ (m_2+m_3)
\left( \frac{f_2^2}{m_2} + \frac{f_3^2}{m_3} \right) -f_1^2 \right] ~~~.
\label{head-on2}
\eeqa

The analysis showed that, if $f^\mu$ is interpreted as a Lorentz force,
the maximal shift of the GZK is at most by a factor of two.
Equation (\ref{head-on2}) is the result of a "back on the envelope
calculation", with
the energy parameter $\omega$ of the photon from the CMB.
It suffices to estimate the shift, but in order to obtain the shape of the
cosmic ray spectrum,
a complete determination involves the folding with the thermal spectrum of
the CMB \cite{berez}.

\section{Conclusions}

In this contribution we presented a pseudo-complex formulation of
Quantum field theory, suggested schematically in Refs.
\cite{schuller3,schuller0}, however, without further calculations.
The pseudo-complex field theory (PCFT) shows important properties, like
i) it contains a maximal acceleration, implying a {\it scalar}
minimal length parameter, due to which ii) it is Pauli-Villars regularized,
iii) maintains the concept of gauge invariance and 
iv) for each particle two mass scales appear, where one is associated
to the usual physical mass and the other one to a mass of Planck scale,
as argued in the main text.

The appearance of a smallest length scale in the theory is by itself
interesting, asking: What are its influences on possible observable
deviations to SFT? Where and how we have to search for it?

A new variational procedure had to be used, leading to the two
mass scales, associated to each particle. The quantization process was
discussed and shown how to define, in a consistent manner, propagators.
Two distinct quantization formulations were investigated. The first one leads
to standard (anti-)commutation relations, while the second one
has an opposite sign in the (anti-)\-commutation relations of the
fields in the $\sigma_-$ component, leading possibly
to a different field theory.
The deep physical consequences of choosing one or
the other are not clear in detail. We indicated that they lead to
some equivalent results, like the field equations, suggesting a
connection.

An extraction procedure has been formulated for obtaining physical
observable numbers, which are pseudo-real.

The cross sections for the scattering of a charged particle at an
external Coulomb field and the Compton scattering
where determined and deviations to SFT calculated. As one result,
differences to SFT are most probably detected, when
processes with fermion internal lines are considered. In this
case, the deviations are proportional to the minimal length scale
$l$, while for photon internal lines the deviation is
proportional to $l^2$. The largest correction is of the order of
$10^{-18}$. These results show that the introduction of a minimal
length scale does not modify sensibly the old results of SFT
at the energies applied up to now.

The effect of $N$ on the effective photon mass was also discussed, leading to
the dispersion relation
$\omega \approx k \left( 1-\frac{1}{2}(lk)^2 \right)$.
At energies of the GZK scale, the corrections are of the order of $10^{-18}$.
At $TeV$ range, these corrections reduce to the order of $10^{-34}$, far
too low to be observed in near future. The actual experimental upper limits
of the correction to the light velocity is of the order of $10^{-16}$,
for the Compton scattering of photons at $50$~TeV \cite{stecker}.

Finally, we discussed a modification of the theory, which
allows a shift of the GZK limit. First results were published in \cite{hess1}.
The relation of the present theory to several geometric approaches was
discussed, showing that there is an overlap, but also differences appear
which have still to be understood. It is important to note that in
the pseudo-complex field theory, as presented here, a clear
method of quantization is available and on how
to extract cross sections. Discussing the geometrical relation we
obtained hints on how to extended the minimal coupling scheme.

Important problems ly ahead:
One has to include the unified electro-weak field theory, because
the interesting processes happen at high energy,
where effects of the $W$ and $Z$ bosons are of significance. It will be
necessary to calculate the dispersion of a proton
when interacting with a photon
of the CMB, producing a pion and $e^\pm$ pairs, in order to obtain the
cross section at ultra high energies. The Auger experiment measures this
cross section. Indeed, the shift of the GKZ limit is presently the only
existing experimental
hint for a new microscopic structure, a smallest length scale.
Another interesting line of investigation is to search for an inclusion
of Gravity in the pseudo-complex formulation, i.e., General Relativity.
The effects of a smallest length scale have to be investigated.
For example, it would be interesting to consider the modified
Scharzschild metric, giving clues on how $l$ affects the structure
of a black hole.

\section*{Acknowledgments}

Financial support from the DFG, DGAPA and CONACyT. P.O.H.
is a Mercator-professor and thanks the FIAS
for all the support and hospitality given. We thank F.P.Schuller for
helpful comments on the pseudo-complex field theory and H. Goenner specially
for helpful comments related to the section of the geometric approach
and for giving us the hint to the work of Crumeyrolle and others.

\appendix

\section{Fourier transforms and the pseudo-complex
Delta function}

Let us start with the pseudo-complex $\delta$-function: It is defined
as

\beqa
\tilde{\delta}(X-Y) & = & \frac{1}{(2\pi )} \int dP e^{iP(X-Y)} ~~~.
\eeqa
A straight line in the "space-like" sector is given by
$X=Rexp(I\phi_0)$, with $\phi_0=const$. The ratio of the
pseudo-imaginary and pseudo-real part is

\beqa
X_1& = & R~{\rm cosh}(\phi_0) \nonumber \\
X_2& = & R~{\rm sinh}(\phi_0) \nonumber \\
\frac{X_2}{X_1} & = & {\rm tanh}(\phi_0) ~~~.
\eeqa
The integration over this straight line gives

\beqa
\tilde{\delta}(X-Y) & = & \frac{1}{(2\pi)} \int dR ~ e^{I\phi_0}
e^{R~e^{I\phi_0}(X-Y)} \nonumber \\
& = & \frac{e^{I\phi_0}}{(2\pi)} \int_{-\infty}^{+\infty}
dR~\left\{ e^{iRe^{\phi_0}(X_+ - Y_+)} \sigma_+  \right. \nonumber \\
& & + \left. e^{iRe^{-\phi_0}(X_- - Y_-)} \sigma_- \right\} \nonumber \\
& = & e^{I\phi_0} \left( \delta (e^{\phi_0}(X_+-Y_+)) \sigma_+  \right. \nonumber \\
& & + \left. \delta (e^{-\phi_0}(X_--Y_-)) \sigma_- \right) ~~~,
\eeqa
where we used that $e^{I\phi_0}=e^{\phi_0} \sigma_+ + e^{-\phi_0} \sigma_-$.
Using this relation again, $\sigma_\pm =0$ and $\sigma_+\sigma_-=0$,
and the property
$\delta(aX)=\frac{1}{|a|}\delta(X)$, we arrive at

\beqa
\delta (X_+-Y_+) \sigma_+ + \delta (X_--Y_-) \sigma_-  ~~~,
\eeqa
which acts like the usual delta function, i.e.

\beqa
\int f(X) \delta(X-Y)~dX  & = & \int f(X_+) \delta(X_+-Y_+) \sigma_+
\nonumber \\
& & + \int f(X_-) \delta(X_--Y_-) \sigma_- \nonumber \\
& = & f(Y_+) \sigma_+ + f(Y_-) \sigma_- \nonumber \\
& = & f(Y) ~~~.
\eeqa

For an integration on a "time-like" straight line, for $X$ the
parametrization $IRexp(I\phi_0)$ has to be used.
The final result is

\beqa
I\left[\delta (X_+-Y_+) \sigma_+ + \delta (X_--Y_-) \sigma_- \right] ~~~,
\eeqa
which acts almost like the usual delta function, i.e.

\beqa
\int f(X) \delta(X-Y)~dX  & = & \int f(X_+) \delta(X_+-Y_+) \sigma_+
\nonumber \\
& & - \int f(X_-) \delta(X_--Y_-) \sigma_- \nonumber \\
& = & f(Y_+) \sigma_+ - f(Y_-) \sigma_- \nonumber \\
& = & I~f(Y) ~~~.
\eeqa
It does not give the function $f(Y)$ but $If(Y)$.

Let us now return to the Fourier transform in 1-dimension. We define it as

\beqa
\tilde{F}(P) & = & \frac{I^{\xi}}{(2\pi )^{\frac{1}{2}}} \int dX F(Y)
e^{-i PY}
~~~.
\eeqa
Let us calculate

\beqa
\frac{1}{(2\pi )^{\frac{1}{2}}} \int dP \tilde{F}(P)
e^{i PX} ~~~.
\eeqa
Substituting $\tilde{F}(P)$, we arrive at

\beqa
& I^{\xi} \int dX ~ \left( \frac{1}{(2\pi)} \int dP~ e^{iP( X-Y )} \right)
F(Y) & ~~~.
\eeqa
The expression in the parenthesis gives $I^{\xi}\delta (X-Y)$, with
$\xi =0$ for an integration along a straight line in the "space-like"
sector and $\xi =1$ for the integration along a straight line in the
"time-like" sector. The final result is $I^{\xi}F(X)$, showing that the
inverse transformation is given by

\beqa
F(X) & = & \frac{I^\xi}{(2\pi )^{\frac{d}{2}}} \int dP \tilde{F}(P)
e^{i PX} ~~~.
\eeqa

For a multi-dimensional integral, the factor in front changes to
$I^{n_2}/(2\pi)^{\frac{d}{2}}$,
where $n_2$ gives the number of times one integrates in a "time-like"
sector and $d$ is the dimension of the integral.
In order to be consistent, one is only allowed to integrate
along straight lines. Arbitrary curves lead to integrals which
can not be related to simple delta-functions.

Wether the extended definition of the $\delta$-function to the pseudo-complex
space plays an important r\^{o}le has still to be verified.

\section{Commutation relations of $a$ and $a^\dagger$}

With a bar above a pseudo-complex variable we indicate usual complex
conjugation. With this we have

\beqa
& \int \bar{f}_P(X) \Phi (X) d^3 X  = 
\int d^3P^\prime \bd{a}(\bd{p}^\prime ) \int \bar{f}_P(X) {f}_{P^\prime}(X)
d^3X & \nonumber \\
& + \int d^3P^\prime \bd{a}^\dagger(\bd{p}^\prime ) \int \bar{f}_P(X)
\bar{f}_{P^\prime}(X)
d^3X &  ~~~.
\eeqa
This involves integrals of the type
$\int e^{i(\bd{P}-\bd{P}^\prime )\cdot X} d^3X$.
In Appendix A we saw that these integrals are related to a generalized
$\tilde{\delta} (\bd{X}-\bd{Y})$ function.

{\it There is, therefore,
an ambiguity over which curve we have to integrate}. This is the same
problem we face for the integration over $X$. In order to
keep a certain liberty, we multiply such an integral with $I^{\xi_P}$,
where $\xi_P$ has yet to be specified. It is zero when the integration
is taken along a straight line in the "space-like"
sector and it is 1 when the
integration path is along a straight line in the "time-like" sector. The
same holds for an integration in the coordinate space, where the index $P$
is substituted by $X$.

Finally we arrive at

\beqa
& \int \bar{f}_P(\bd{X},x_0) \Phi (\bd{X},X_0)  = & \nonumber \\
& I^{\xi_X}\left\{ \frac{1}{2\omega_P} \bd{a}(\bd{P}) + \frac{1}{2\omega_P}
\bd{a}^\dagger (-\bd{P}) e^{2i\omega_P X_0}  \right\}  &    ~~~.
\eeqa

Using the conjugate momentum of the field, we get also

\beqa
& \int \bar{f}_P(\bd{X},x_0) D_0\Phi (\bd{X},X_0)  = &  \nonumber \\
& I^{\xi_X}\left\{ -\frac{i}{2} \bd{a}(\bd{P}) + \frac{i}{2}
\bd{a}^\dagger (-\bd{P}) e^{2i\omega_P X_0} \right\}   &     ~~~.
\eeqa

The two equations lead to

\beqa
\bd{a}(\bd{P}) & = & iI^{\xi_X} \int d^3X \bar{f}_P(\bd{X},X_0)
\leftrightarrow{D}_0\Phi (\bd{X},X_0) \nonumber \\
\bd{a}^\dagger (\bd{P}) & = & -iI^{\xi_X} \int d^3X f_P(\bd{X},X_0)
\leftrightarrow{D}_0\Phi (\bd{X},X_0) ~~~,
\eeqa
with $A(X_0)\leftrightarrow{D}_0B(X_0)$ = $A(X_0)\left( D_0B(X_0) \right) -
\left( D_0 A(X_0) \right) B(X_0)$.

The commutator of the operators $\bd{a}$ and $\bd{a}^\dagger$ is,
using Eq. (\ref{phi-pi}),

\beqa
\left[ \bd{a}(\bd{P}),\bd{a}^\dagger (\bd{P}^\prime ) \right] & = &
I^{n+ \xi_X} \delta^{(3)}(\bd{P} - \bd{P}^\prime) ~~~,
\label{a-a-dagger}
\eeqa
where one integration in $X$ had to be performed.

It depends now very much on the path of integration, which quantization
property we obtain of the operators $\bd{a}$ and $\bd{a}^\dagger$.
This might have also consequences on the physical theory, which are
not clear yet.

Conversely, starting from the commutation relation (\ref{a-a-dagger})
of the creation and annihilation operators, the one for the field and
its conjugate momentum is

\beqa
\left[ \Phi (\bd{X},X_0), \Pi(\bd{Y},X_0) \right] & = &
iI^{n+\xi_x + \xi_p}\delta^{(3)} (\bd{X}-\bd{Y}) ~~~,
\eeqa
which requires to set $\xi_x + \xi_p$ to even values, for reasons of
consistency to Eq. (\ref{phi-pi}). This implies that either both are
zero or both are 1. It is not clear what an integration along the
pseudo-imaginary axis of the momentum means, i.e., the case
$\xi_x=\xi_p=1$..


\begin{thebibliography}{99}

\bibitem{greiner} W. Greiner and J. Reinhardt, {\it Quantum Electrodynamics},
3rd ed. (Springer, Heidelberg, 2003).

\bibitem{agasa} M Takeda et al., {\it Phys. Rev. Lett.} {\bf 81}, 1163 (1998).


\bibitem{fly}  D. J. Bird et al., {\it Astrophys. J.} {\bf 441}, 144 (1995);
Astrophys. J. {\bf 424}, 491 (1994); Phys. Rev. Lett. {\bf 71}, 3401 (1993).

\bibitem{haverah} M. A. Lawrence, R. J. O. Reid and A. A. Watson,
{\it J. Phys. G} {\bf 17}, 773 (1991).

\bibitem{yakutsk} N. N. Efimov et al., {\it Proceedings of the
International Symposium on Astrophysical Aspects of the Most Energetic
Cosmic Rays}, Eds. M. Nagano and F. Takahara, (World Scientific, Singapore,
1991), p. 20.

\bibitem{gkz1} K. Greisen, {\it Phys. Rev. Lett.} {\bf 16}, 748 (1966).

\bibitem{gkz2} G. T. Zatsepin and V. A. Kuzmin, {\it JETP Lett.} {\bf 4}, 78 
(1966).

\bibitem{HiRes} R. U. Abbasi et al., astro-ph/0703099, (2007).

\bibitem{auger} J. W. Cronin, {\it Nucl. Phys. B (Proc. Supl.)} {\bf 28}, 78
(1992).

\bibitem{lukas} The Pierre Auger Collaboration, astro-ph/0606619.

\bibitem{dsr1} G. Amelino-Camelia and T. Piran, {\it Phys. Rev. D} {\bf 64}, 036005
(2001).

\bibitem{dsr2} G. Amelino-Camelia, {\it Int. J. Mod. Phys. D} {\bf 11}, 35
(2002).

\bibitem{smolin} R. Borissov, S. Major and L. Smolin, {\it Class. Quant. Grav.}
{\bf 13}, 3183 (1996).

\bibitem{goenner1} G. Yu. Bogoslovsky and H. F. Goenner,
{\it Phys. Lett. A} {\bf 244}, 222 (1998).

\bibitem{goenner2}  H. F. Goenner and G. Yu. Bogoslovsky,
{\it Gen. Rel. and Grav.} {\bf 31}, 1383 (1999).

\bibitem{goenner3} G. Yu. Bogoslovsky and H. F. Goenner,
{\it Gen. Rel. and Grav.} {\bf 31}, 1565 (1999).

\bibitem{bertolami} O. Bertolami and J. G. Rosa, {\it Phys. Rev. C} {\bf 71}, 097901
(2005).

\bibitem{coleman1} S. Coleman and S. L. Glashow, {\it Physics Letters
B} {\bf 405}, 249 (1997).

\bibitem{coleman2} S. Coleman and S. L. Glashow, {\it Phys. Rev. D} {\bf 59}, 116008
(1999).

\bibitem{hess1} P. O. Hess and W. Greiner, submitted for publication (2007).

\bibitem{schuller2} F. P. Schuller, {\it Ann. Phys. (N.Y.)} {\bf 299}, 174 (2002).

\bibitem{crumeyrolle} A. Crumeyrolle, {\it Annales de la Facult\'e de Sciences,
Toulouse} {\bf 26}, 105 (1964).

\bibitem{schuller3} F. P. Schuller, M. N. R. Wohlfarth and T. W. Grimm,
{\it Class. Quant. Grav.} {\bf 20}, 4269 (2003).

\bibitem{schuller1} F. P. Schuller, {\it Phys. Lett. B} {\bf 540}, 119 (2002).

\bibitem{schuller4} F. P. Schuller, {\it Eur. Phys. J. C} {\bf 39} (S3), 13
(2004).

\bibitem{kantor} I. L. Kantor, A. S. Solodovnikov, {\it Hypercomplex
Numbers. An Elementary Introduction to Algebra}, (Springer, Heidelberg,1989).

\bibitem{schuller0} F. P. Schuller, Ph.D. thesis, University of Cambridge
(2003).

\bibitem{peschl} E. Peschl, {\it Funktionentheorie},
(BI-Hochschultaschenb\"ucher, Band 131, Mannheim, 1967).

\bibitem{greiner-cl} W. Greiner, {\it Classical Mechanics} I and II,
3rd edition, (Springet, Heidelberg, 1989).

\bibitem{gen-lo} R. Sexl, H. K. Urbanke, {\it Relativit\"at, Gruppen,
Teilchen}, (Springer, Heidelberg, 1976).

\bibitem{mueller} W. Greiner and B. M\"uller, {\it Quantum Mechanics:
Symmetries}, (Springer, Heidelberg, 1994).

\bibitem{greiner2} W. Greiner, J. Reinhardt, {\it Field Quantization},
(Springer, Heidelberg, 1996).

\bibitem{stumpf} H. Stumpf, {\it Z. Naturforschung} {\bf 55a}, 415 (2000).

\bibitem{stecker} F. W. Stecker, Astropart. Phys. {\bf 20}, 85 (2003).

\bibitem{caneloni} E. R. Caianiello, {\it Nuovo Cim. Lett.} {\bf 32}, 65 (1981).

\bibitem{born1} M. Born, {\it Proc. Roy. Soc. A} {\bf 165}, 291 (1938).

\bibitem{born2} M. Born, {\it Rev. Mod. Phys.} {\bf 21}, 463 (1949).

\bibitem{low1} S. G. Low, {\it J. Math. Phys.} {\bf 38}, 2197 (1997).

\bibitem{beil1} R. G. Beil, {\it Found. Phys.} {\bf 33}, 1107 (2003).

\bibitem{tach} S. Tachibana and M. Okumura, {\it Tohoku Math. J.}
{\bf 14}, 156 (1962). 

\bibitem{low2} S. G. Low, {\it J. Phys. A} {\bf 35}, 5711 (2002).

\bibitem{beil2} R. G. Beil, {\it Int. J. Theor. Phys.} {\bf 26}, 189 (1987).

\bibitem{beil3} R. G. Beil, {\it Int. J. Theor. Phys.} {\bf 28}, 659 (1989).

\bibitem{beil4} R. G. Beil, {\it Int. J. Theor. Phys.} {\bf 31}, 1025 (1992).

\bibitem{goenner4} H. F. Goenner, private communication (2007).

\bibitem{brandt1} H. E. Brandt, {\it Found. Phys. Lett.} {\bf 2}, 39 (1989).

\bibitem{brandt2} H. E. Brandt, {\it Found. Phys. Lett.} {\bf 4}, 523 (1991).

\bibitem{brandt3} H. E. Brandt, {\it Found. Phys. Lett.} {\bf 6}, 245 (1993).

\bibitem{crum2} A. Crumeyrolle, {\it Riv. Mat. Univ. Parma} {\bf 5}, 85 (1964).

\bibitem{crum3} A. Crumeyrolle, {\it Ann. de L'I. H. P., Section A} {\bf 29}, 217
(1978).

\bibitem{moffat1} J. W. Moffat, {\it Phys. Rev. D} {\bf 19}, 3554 (1979).

\bibitem{kunstatter} G. Kunstatter and R. Yates, {\it J. Phys. A}
{\bf 14}, 847 (1981).

\bibitem{moffat2} G. Kunstatter, J. W. Moffat and J. Malzan,
{\it J. Math. Phys.} {\bf 24}, 886 (1983).

\bibitem{mann} R. B. Mann, {\it Class. Quant. Grav.} {\bf 1}, 561
(1984).

\bibitem{taizo} T. Muta, {\it Foundations of Quantum Chromo Dynamics},
(World Scientific, Singapore, 1987).

\bibitem{berez} V. S. Berezinsky and S. I. Grigor'eva, {\it Astron. Astrophys.}, 1
(1988).

\end{thebibliography}
\end{document}